
\input jytex.tex   
\typesize=10pt
\magnification=1200
\baselineskip=17truept
\hsize=6truein\vsize=8.5truein
\sectionnumstyle{blank}
\chapternumstyle{blank}
\chapternum=1
\sectionnum=1
\pagenum=0

\def\begintitle{\pagenumstyle{blank}\parindent=0pt\begin{narrow}[0.4in]}
\def\endtitle{\end{narrow}\newpage\pagenumstyle{arabic}}


\def\beginexercise{\vskip 20truept\parindent=0pt\begin{narrow}[10
truept]}
\def\endexercise{\vskip 10truept\end{narrow}}


\def\eql#1{\eqno\eqnlabel{#1}}
\def\ref{\reference}
\def\peq{\puteqn}
\def\pref{\putref}

\def\mgn{\marginnote}
\def\bex{\begin{exercise}}
\def\eex{\end{exercise}}

\font\open=msbm10 
\def\mbox#1{{\leavevmode\hbox{#1}}}
\def\hspace#1{{\phantom{\mbox#1}}}
\def\oZ{\mbox{\open\char90}}

\def\rB{{\rm B}}

\def\al{\alpha}
\def\be{\beta}
\def\ga{\gamma}
\def\de{\delta}
\def\Ga{\Gamma}

\def\ep{\epsilon}

\def\la{\lambda}

\def\si{\sigma}

\def\ze{\zeta}

\def\De{\Delta}

\def\Real{{\rm Re\,}}

\def\zf{$\zeta$--function}
\def\zfs{$\zeta$--functions}


\def\frac#1/#2{\leavevmode\kern.1em
\raise.5ex\hbox{\the\scriptfont0 #1}\kern-.1em/\kern-.15em
\lower.25ex\hbox{\the\scriptfont0 #2}}
\def\sfrac#1/#2{\leavevmode\kern.1em
\raise.5ex\hbox{\the\scriptscriptfont0 #1}\kern-.1em/\kern-.15em
\lower.25ex\hbox{\the\scriptscriptfont0 #2}}

\def\gtorder{\mathrel{\raise.3ex\hbox{$>$}\mkern-14mu
             \lower0.6ex\hbox{$\sim$}}}
\def\ltorder{\mathrel{\raise.3ex\hbox{$<$}\mkern-14mu
             \lower0.6ex\hbox{$\sim$}}}

\def\semidirprod{\rlap{\ss C}\raise1pt\hbox{$\mkern.75mu\times$}}
\def\for{\lower6pt\hbox{$\Big|$}}
\def\fish{\kern-.25em{\phantom{abcde}\over \phantom{abcde}}\kern-.25em}


\def\boxit#1{\vbox{\hrule\hbox{\vrule\kern3pt
        \vbox{\kern3pt#1\kern3pt}\kern3pt\vrule}\hrule}}
\def\dalemb#1#2{{\vbox{\hrule height .#2pt
        \hbox{\vrule width.#2pt height#1pt \kern#1pt
                \vrule width.#2pt}
        \hrule height.#2pt}}}


\def\noin{\noindent}

\def\comb#1#2{{\left(#1\atop#2\right)}}

\def\eg{{\it e.g. }}
\def\ie{{\it i.e. }}
\def\cf{{\it cf }}
\def\pa{\partial}


  %

\def\sumstart#1#2{{\mathop{{\sum}^*}_{#1}^{#2}}}
\def\sumstar#1{{\mathop{{\sum}^*_{#1}}}}

\def\3j#1#2#3#4#5#6{\left\lgroup\matrix{#1&#2&#3\cr#4&#5&#6\cr}
\right\rgroup}

\def\man{{\cal M}}

\def\m?{\mgn{?}}


\def\aop#1#2#3{{\it Ann. Phys.} {\bf {#1}} (19{#2}) #3}

\def\cmp#1#2#3{{\it Comm. Math. Phys.} {\bf {#1}} (19{#2}) #3}
\def\cqg#1#2#3{{\it Class. Quant. Grav.} {\bf {#1}} (19{#2}) #3}

\def\jmp#1#2#3{{\it J. Math. Phys.} {\bf {#1}} (19{#2}) #3}
\def\jpa#1#2#3{{\it J. Phys.} {\bf A{#1}} (19{#2}) #3}

\def\np#1#2#3{{\it Nucl. Phys.} {\bf B{#1}} (19{#2}) #3}
\def\pl#1#2#3{{\it Phys. Lett.} {\bf {#1}} (19{#2}) #3}
\def\pm#1#2#3{{\it Phil.Mag.} {\bf {#1}} ({#2}) #3}

\def\pr#1#2#3{{\it Phys. Rev.} {\bf {#1}} (19{#2}) #3}
\def\prA#1#2#3{{\it Phys. Rev.} {\bf A{#1}} (19{#2}) #3}

\def\prD#1#2#3{{\it Phys. Rev.} {\bf D{#1}} (19{#2}) #3}

\def\prs#1#2#3{{\it Proc. Roy. Soc.} {\bf A{#1}} (19{#2}) #3}
\def\pcps#1#2#3{{\it Proc. Camb. Phil. Soc.} {\bf{#1}} (19{#2}) #3}

\def\dmj#1#2#3{{\it Duke Math. J.} {\bf {#1}} (19{#2}) #3}

\def\jdg#1#2#3{{\it J. Diff. Geom.} {\bf {#1}} (19{#2}) #3}
\def\jfa#1#2#3{{\it J. Func. Anal.} {\bf {#1}} (19{#2}) #3}

\def\ma#1#2#3{{\it Math. Ann.} {\bf {#1}} ({#2}) #3}
\def\mz#1#2#3{{\it Math. Zeit.} {\bf {#1}} ({#2}) #3}
\def\pams#1#2#3{{\it Proc. Am. Math. Soc.} {\bf {#1}} (19{#2}) #3}

\def\qjm#1#2#3{{\it Quart. J. Math.} {\bf {#1}} (19{#2}) #3}

\def\tams#1#2#3{{\it Trans. Am. Math. Soc.} {\bf {#1}} (19{#2}) #3}

\begin{title}
\vglue 20truept
\righttext {MUTP/95/7}
\righttext{hep-th/9506042}
\leftline{\today}
\vskip 100truept
\centertext {\Bigfonts \bf Robin conditions on the Euclidean ball}
\vskip 15truept
\centertext{J.S.Dowker\footnote{Dowker@a3.ph.man.ac.uk}}
\vskip 7truept
\centertext{\it Department of Theoretical Physics,\\
The University of Manchester, Manchester, England.}
\vskip 60truept
\centertext {Abstract}
\begin{narrow}
Techniques are presented for calculating directly the scalar functional
determinant on the Euclidean $d$-ball. General formulae are given for
Dirichlet and Robin boundary conditions. The method involves a large mass
asymptotic limit which is carried out in detail for $d=2$ and $d=4$
incidentally producing some specific summations and identities. Extensive
use is made of the Watson-Kober summation formula.
\end{narrow}
\vskip 5truept
\righttext {June 1995}
\vskip 75truept
\righttext{Typeset in \jyTeX}
\vfil
\end{title}
\pagenum=0
\section{\bf 1. Introduction}
For specially symmetric situations it is often possible to relate the
eigenvalue problem on a bounded region, $\man$, to one on a covering space,
the method of images being an example. As Pockels [\pref{Pockels}] has
pointed out, in the case of the boundary condition,
$\big({\bf n.}\nabla\phi-h\phi\big)\big|_{\pa\man}=0$, this technique is
no longer applicable and one has to proceed directly.
In classical physics such conditions arise in heat conduction
[\pref{CandJ}].

In the present work we wish to analyse some cases of these Robin
boundary
conditions, evaluating the coefficients in the asymptotic expansion of the
heat-kernel and also the corresponding functional determinant. The reason
is that it is often useful to have explicit results for special cases in
order to put restrictions on the general situation. In addition, Robin
conditions enter into the mixed boundary conditions required for higher
dimensional vector bundles.

The domain of concern here is the Euclidean $d$-ball. The determinants
on such regions have already been found, in the massless case, by a
conformal transformation
method but this technique is limited, at the moment, to $d\le4$ by the
availability of the necessary transformation law. A more direct method is
therefore needed.

The recent determination by Branson, Gilkey and Vassilevich [\pref{BGV}]
of the $C_{5/2}$ coefficient is not sufficient to allow the extension
to $d=5$ as it is restricted to totally geodesic boundaries.

Other calculations involving relevant material that should be mentioned
are those by D'Eath and Esposito [\pref{DandE,Esposito}], Schleich
[\pref{Schleich}] and Louko [\pref{Louko}]. A parallel calculation by
Bordag {\it et al} has recently appeared, [\pref{BGKE}].

\section {\bf 2. The interval}
We work on $I=[0,1]$ and consider the simplest equation
$$-{d^2\phi\over dx^2}=\la\phi$$
with
$${d\phi\over dx}\bigg|_1=-h_1\phi_1,\quad
{d\phi\over dx}\bigg|_0=h_0\phi_0.$$

This eigenvalue problem is discussed in Carslaw and Jaeger. We write the
eigenfunctions
$$\phi(x)=N\sin(\al x+\de)
$$ where $\la=\al^2$ and $\tan\de=\al/h_0$. The eigenvalue equation is
derived to be
$${\tan\al\over\al}={h_1+h_0\over\al^2-h_1h_0}\,.
\eql{eigv1}$$
The normalisation, $N$ is easily worked out.

If $h_0=h_1=h$, (\peq{eigv1}) factorises into
$$\bigg(2h\cos(\al/2)+\al\sin(\al/2)\bigg)
\bigg(h{\sin(\al/2)\over\al}-2\cos(\al/2)\bigg)=0.
\eql{eigv2}$$

An eigenvalue equation similar to (\peq{eigv1}), but with the right-hand
side reversed in sign, occurs in the theory of the
three-ball with Neumann conditions at the surface (Rayleigh
[\pref{Rayleigh}] I.p265, eqn.(2)).

The eigenvalues are non-degenerate and, although easily computed,
are not needed here since the Mittag-Leffler theorem can be applied
$$(h_0h_1-z^2){\sin z\over z}+(h_1+h_0)\cos z=
\big(h_0h_1+h_0+h_1\big)
\prod_{\al}\bigg(1-{z^2\over\al^2}\bigg)
\eql{mittag}$$
where the product is over the positive roots.
\begin{ignore}
the Mittag-Leffler theorem can be applied to each of the factors

$$\eqalign{
\bigg((h_1+h_0)\cos(z/2)+\al\sin(z/2)\bigg) &=(h_0+h_1)
\prod_{\al_0}\bigg(1-{z^2\over\al_0^2}\bigg)\cr
\bigg(h_1h_0{\sin(z/2)\over z}-(h_1+h_0)\cos(z/2)\bigg)
&=({h_1h_0\over2}-h_1-h_0)\prod_{\al_1}\bigg(1-
{z^2\over\al_1^2}\bigg).\cr}
\eql{mittag}$$
and the products are over the positive roots.
\end{ignore}

Incidentally, logarithmic differentiation, expansion in positive powers
of $z$ and the equating of coefficients allows one to evaluate
the sums of inverse powers of the roots. (This technique goes back to
Euler, \cf Rayleigh [\pref{Rayleigh}] II p.279. Rayleigh does not
differentiate but just expands the logarithms. See also Watson
[\pref{Watson1}] 15.5, 15.51.)

It is found that, for example,
$$\eqalign{
&\sum_{\al}{1\over\al^2}={h_0h_1+3(h_0+h_1)+6\over6(h_0h_1+h_0+h_1)}\,,\cr
&\sum_{\al}{1\over\al^4}=
{h_0^2h_1^2+6(h_0^2h_1+h_0h_1^2)+15(h_0^2+h_1^2)+30h_0h_1+60(h_0+h_1)+90
\over90(h_0h_1+h_0+h_1)^2}\cr}
$$
which can be confirmed numerically.
\begin{ignore}
$$\eqalign{
&\sum_{\al_0}{1\over\al_0^2}={h_0+4\over4h_0};
\quad\sum_{\al_0}{1\over\al_0^4}={h_0^2+8h_0+24\over6h_0^2}\cr
&\sum_{\al_1}{1\over\al_1^2}={h_1+6\over24(h_1+2)};
\quad\sum_{\al_1}{1\over\al_1^4}={h_1^2+12h_1+60\over1440(h_1+2)^2}\cr}
\eql{rootsums}$$
\end{ignore}
Defining the \zf\ by
$$\ze(s)=\sum_\la{1\over\la^s}
$$
we see that the values $\ze(n)$, $n\in\oZ^+$, can be determined.

One of our aims is to find the coefficients in the heat-kernel expansion.
These are related to the residues at the poles of $\ze(s)$ and to the
values
$\ze(-n)$, $n\in\oZ^+$. The method we use here involves the time-honoured
one of the asymptotic behaviour of the resolvent
$$
R(m^2)=\sum_\la{1\over\la+m^2}
\eql{re}$$
as $m^2\to\infty$. (\cf Dikii,[\pref{Dikii}], and [\pref{Voros,Moss}]).

More generally we can define
$$\ze(s,m^2)=\sum_\la{1\over(\la+m^2)^s},
\eql{zem}$$
and it is straightforward to derive the large, positive $m$
asymptotic expansions
(\cf Dowker and Critchley [\pref{DandC1}], Voros [\pref{Voros}], Elizalde
[\pref{Elizalde1}]),
$$\ze(s,m^2)\sim{1\over(4\pi)^{d/2}}\sum_{n=0,1/2,1,\ldots}
(m^2)^{d/2-n-s}{\Ga(s-d/2+n)\over\Ga(s)}\,C_n.
\eql{asympm1}$$ and
$$\ze'(0,m^2)\sim{1\over(4\pi)^{1/2}}\sum_{n=0,1,3/2,\ldots}\bigg(C_n
\Ga(n-1/2)m^{-2n+1}-C_{1/2}\ln m^2\bigg).
\eql{asympm2}$$

If $d=1$ it is sufficient to set $s=1$ in (\peq{asympm1}) to get
$$
R(m^2)\sim{1\over(4\pi)^{1/2}}\sum_{n=0,1/2,1,\ldots}
{\Ga(1/2+n)\over(m^2)^{1/2+n}}\,C_n.
\eql{resm}$$

Comparison with (\peq{mittag}) shows that $z=im$ and that
$$\eqalign{
R(m^2)&={d\over dm^2}\ln\bigg((h_0h_1+m^2){\sinh m\over m}
+(h_1+h_0)\cosh m\bigg).\cr}
\eql{resolv}$$

The large positive $m$ limit is easy to find,
$$\eqalign{
R(m^2)&\sim {1\over2m}\bigg(1+{1\over m+h_1}+ {1\over m+h_0}-{1\over
m}\bigg)\cr
&={1\over2m}\bigg(1+{1\over m}+\sum_{k=1}^\infty(-1)^k\,{h_1^k+h_0^k\over
m^{k+1}}\bigg)\cr}
$$
and the coefficients $C_n$ can be read off as
$$\eqalign{
&C_0=1,\cr
&C_{1/2}=\sqrt\pi,\cr
&C_n= -{(-1)^{2n}\sqrt\pi\over\Ga(n+1/2)}\,(h_1^{2n-1}+h_0^{2n-1})\,;
\quad n\ge 1. \cr}
\eql{coeffs}$$
These check with known forms and we are now in possession of the
values $\ze(-n)$.

Our next objective is the functional determinant,
$$D(-m^2)=\exp\big(-\ze'(0,m^2)\big)
\eql{det}$$ in particular $D(0)$.

{}From (\peq{mittag})
$$
\ze'(0,0)-\ze'(0,m^2)=\ln F(m^2)
-\ln\big(h_0h_1+h_0+h_1\big),
\eql{zedashdiff}$$
where $F$ is the factor in (\peq{resolv}).
We can thus write
$$
\ze'(0,0)=\lim_{m\to\infty}\bigg(\ze'(0,m^2)+\ln F(m^2)
-\ln\big(h_0h_1+h_0+h_1\big)\bigg).
$$

{}From (\peq{resolv})
$$
\ln F(m^2)\to m+\ln m-\ln2+
\ln\bigg(\big(1+{h_0\over m}\big)\big(1+{h_1\over m}\big)\bigg),
\eql{asymp3}$$
and using the asymptotic form (\peq{asympm2}) one finds that
$$
\ze'(0,0)=-\ln2-\ln\big(h_0h_1+h_1+h_0\big)
\eql{zedash}$$
and so $D(0)=2\big(h_0h_1+h_1+h_0\big)$. The same result emerges from the
method of Barvinsky {\it et al}, [\pref{Barv}].

As a check, expanding the logarithm in (\peq{asymp3}) gives the previous
expressions, (\peq{coeffs}), for the coefficients.

The Neumann (and Dirichlet) value for $\ze'(0,0)$ is $-\ln2$. But the
Neumann case, $h_1=h_0=0$, has to be treated afresh since there are
zero modes. The vanishing of the determinant, \ie $h_0h_1+h_1+h_0=0$,
is related to the appearance of a zero eigenvalue but the corresponding
mode does not exist.

Turning equation (\peq{zedashdiff}) around, one gets the massive functional
determinant,
$$
D(-m^2)=2\bigg((h_0h_1+m^2){\sinh m\over m}+(h_1+h_0)\cosh m\bigg).
\eql{mfuncdet1}$$

Other, miscellaneous boundary conditions, [\pref{CandJ}], can be treated
in the same way with similar results.
\begin{ignore}
It is convenient to represent the \zf\ by
a contour integral. From the product form (\peq{mittag}) it follows that
$$
\ze(s)={1\over2\pi i}\int_\Ga {d\over dz^2}\ln\big(F_0(z^2)F_1(z^2)\big)\,
{dz^2\over z^{2s}}.
\eql{int1}$$
The contour $\Ga$
encloses all the eigenvalues $\al^2$ and can be deformed to run around the
cut along the negative $z^2$ axis. This yields, for $\Re s>1/2$,
$$\eqalign{
\ze(s)={\sin\pi s\over\pi}
\int_\ep^\infty{d\over dm^2}&\ln\big(F_0(-m^2)F_1(-m^2)\big)\,
{dm^2\over m^{2s}}\cr
&+{1\over2\pi i}\int{d\over dz^2}\ln\big(F_0(z^2)F_1(z^2)\big)\,
{dz^2\over z^{2s}}\cr}
$$ where the loop integral is over a circle of radius $\ep$ centred
on the origin,
\cf [\pref{Barv,Dikii}]. The second integral tends to zero
as $\ep\to0$ for $1/2<\Re s<1$ and the first integral remains finite.
\end{ignore}

\section{\bf 3. Heat kernel coefficients on Euclidean balls.}
The higher dimensional interval, \ie the ball, is treated by the same
approach. The operator is the Laplacian
and the standard mode expressions, which will not be repeated here, are
outlined by Moss [\pref{Moss}]. (See also [\pref{MandO,Erdelyi}].)
The eigenvalues $\la=\al^2$ in (\peq{zem}) are given by the roots of
$$\eqalign{
J_p(\al)&=0,\quad{\rm Dirichlet}\cr
\be J_p(\al)+&\al J_p'(\al)=0,\quad{\rm Robin},\cr}
\eql{roots}$$
where $\be=h+1-d/2$. The degeneracy for a given Bessel order
is, for even $d$
$$N^{(d)}_p={2\over(d-2)!}p^2(p^2-1)\ldots\big(p^2-(d/2-2)^2\big),
\eql{degene}$$
while for odd $d$,
$$N^{(d)}_p={2\over(d-2)!}p(p^2-1/4)\ldots\big(p^2-(d/2-2)^2\big).
\eql{degeno}$$

The Mittag-Leffler theorem can again be employed. Since there are no
normalisable zero modes, the zero roots of (\peq{roots}) must be removed.
We therefore write
$$
z^{-p}F_p(z)=\ga\prod_\al\bigg(1-{z^2\over\al^2}\bigg)
\eql{ml1}$$where
$$\eqalign{
\ga&={1\over2^pp!},\quad{\rm Dirichlet}\cr
&={(\be+p)\over 2^pp!},\quad{\rm Robin}\quad(\be,p)\ne(0,0).\cr}
$$
If $\be=0$, then $p=0$ is a special case, $F_0=zJ'_0=-zJ_1$, and
$$
z^{-2}F_0(z)=-{1\over2}\prod_\al\bigg(1-{z^2\over\al^2}\bigg).
\eql{ml3}$$

Moss [\pref{Moss}] was concerned to evaluate the heat-kernel coefficients,
and we present
here an equivalent treatment. The coefficients can be read off from the
asymptotic expansion (\peq{asympm1}) with $s=(d+2)/2$ if $d$ is even and
$s=(d+1)/2$ if $d$ is odd. The requisite expression for the left-hand side
can be obtained by repeated differentiation of the resolvent (\peq{re}) or,
equivalently, of the product (\peq{ml1}), which is rewritten with $z=im$ as
$$
m^{-p}F_p(m)=\ga\prod_\al\bigg(1+{m^2\over\al^2}\bigg)
\eql{ml2}$$where $F_p$ is now given by
$$\eqalign{
F_p(m)&=I_p(m),\quad{\rm Dirichlet}\cr
&=\be I_p(m)+m I_p'(m),\quad{\rm Robin},\cr}
\eql{eff}$$ in terms of the modified Bessel function. (See also D'Eath and
Esposito [\pref{DandE,Esposito}].)

The required quantity is then (\cf [\pref{Moss}])
$$
\ze(q,m^2)=-{1\over(q-1)!}\bigg(-{\pa\over\pa m^2}\bigg)^q
\sum_pN_p\ln\big(m^{-p}F_p(m)\big).
\eql{lhs}$$

The development of the logarithm follows from Olver's asymptotic expansion
of the Bessel functions. In particular
$$
\ln\big(m^{-p}I_p(m)\big)\sim-{1\over2}\ln(2\pi)+\ep-p\ln\big(p+\ep\big)
-{1\over2}\ln\ep+\ln\bigg(1+\sum_{n=1}^\infty{U_n(t)\over p^n}\bigg)
\eql{logex1}$$
and
$$
\ln\big(m^{-p+1}I'_p(m)\big)\sim-{1\over2}\ln(2\pi)+\ep-p\ln\big(p+\ep\big)
+{1\over2}\ln\ep+\ln\bigg(1+\sum_{n=1}^\infty{V_n(t)\over p^n}\bigg)
\eql{logex2}$$
where $\ep=\sqrt(p^2+m^2)$ and $t=p/\ep$. It is sometimes more convenient
to use Olver's variable, $z=m/\ep$, and we will use the same kernel
symbols
for the polynomials whether they are written as functions of $t$ or of $z$.

The polynomials $U_n(t)$ satisfy the recursion relation [\pref{Olver}]
$$
U_{n+1}={1\over2}t^2(1-t^2){dU_n\over dt}+{1\over8}\int_0^t(1-5t^2)\,
U_n dt,
\eql{rec}$$
and the $V_n$ are related to the $U_n$ by
$$
V_n=U_n-{1\over2}t(1-t^2)\big(U_{n-1}+2t{dU_{n-1}\over dt}\big).
$$

Equation (\peq{logex1}) is relevant for the Dirichlet case. The Robin
expansion is
$$
\ln\big(m^{-p}F_p\big)\sim\ep-p\ln\big(p+\ep\big)-{1\over2}\ln(2\pi)
+{1\over2}\ln\ep+\ln\bigg(1+\sum_{n=1}^\infty{W_n\over p^n}\bigg),
\eql{logex3}$$
where $W_n=V_n+\be tU_{n-1}$ with $U_0=1$.

{}From the recursion (\peq{rec}) it is clear that $U_n(t)$, $V_n(t)$ and
$W_n(t)$ each have a factor of $t^n$. Extracting this factor, we write \eg
$$
{U_n(t)\over p^n}={\overline U_n(t)\over\ep^n},
$$
so that  $\overline W_n=\overline V_n+\be \overline U_{n-1}$

The systematic treatment of the summations in (\peq{logex1}) and
(\peq{logex3}) involves the cumulant, or cluster, expansion, written
generally as
$$
\ln\bigg(1+\sum_{n=1}^\infty{Q_n(t)\over\ep^n}\bigg)
=\sum_{n=1}^\infty{S_n(t)\over \ep^n}
$$with
$$S_n=\sum_{\{l\}}(-1)^{\de-1}(\de-1)!\prod_{j=1}^n {Q_j^{l_j}\over l_j!},
\eql{cumul}$$
where $\de=\sum_{j=1}^n l_j$ and the summation is over sets of positive
integers $l_j$ including zero such that $\sum_{j=1}^n j\,l_j=n$ .

For example,
$$\eqalign{
&S_1=Q_1\cr
&S_2=Q_2-{1\over2}Q_1^2\cr
&S_3=Q_3-Q_2Q_1+{1\over3}Q_1^3\cr
&S_4=Q_4-Q_3Q_1-{1\over2}Q_2^2+Q_2Q_1^2-{1\over4}Q_1^4.\cr
&S_5=Q_5-Q_4Q_1-Q_3Q_2+Q_3Q_1^2-Q_2Q_1^3+Q_2^2Q_1+{1\over5}Q_1^5\cr}
$$
the number of terms increasing rapidly with the order $n$.

Specific evaluation shows agreement with ref. [\pref{Moss}], \eg in the
Dirichlet case, (\peq{logex1})
$$\eqalign{
T_1(t)&=-{1\over12}+{5\over24}z^2,\cr
T_2(t)&=-{1\over4}z^2+{5\over16}z^4,\cr
T_3(t)&={1\over360}+{61\over240}z^2-{221\over192}z^4+{1105\over1152}z^6\cr
T_4(t)&=-{1\over4}z^2+3z^4-{113\over16}z^6+{565\over128}z^8\cr
T_5(t)&=-{1\over1260}+{125\over504}z^2-
{493\over72}z^4+{1621\over48}z^6-{82825\over1536}z^8+{82825\over3072}
z^{10}\cr}
\eql{dtees}$$
while, for Robin conditions, the corresponding polynomials are
$$\eqalign{
R_1(\be,t)&=\be-{1\over12}-{7\over24}z^2,\cr
R_2(\be,t)&=-{1\over2}\be^2+{1\over4}(2\be+1)z^2-{7\over16}z^4\cr
R_3(\be,t)&={1\over360}+{1\over3}\be^3-{59\over240}z^2-{1\over2}\be^2z^2
+{259\over192}z^4+{7\over8}\be z^4-{1463\over1152}z^6.\cr}
\eql{rtees}$$

Related expansions for the spin-half case were computed by D'Eath and Esposito
[\pref{DandE,Esposito}].

The next step is the differentiation in (\peq{lhs}) followed by
the sum over $p$, which is the hardest section of the analysis.

To check the method, and for illustrative purposes, we firstly look at
the Dirichlet 2-disc, $q=2$ and $N^{(2)}_p=2-\de_{p0}$ in (\peq{lhs}).
Algebra produces
$$
\bigg({d\over dm^2}\bigg)^2\ln\big(m^{-p}I_p\big)\sim
{1\over2m^2}\bigg({p-\ep\over m^2}+{1\over2\ep}\bigg)+{1\over4\ep^4}+
\bigg({d\over dm^2}\bigg)^2\sum_{n=1}^\infty{T_n(p/\ep)\over \ep^n}
\eql{disc2}$$
for $p>0$.

Since the bracketed term in (\peq{disc2}) converges overall as $p\to
\infty$ it is legitimate to regularise it. Moss writes
$$
\bigg({p-\ep\over m^2}+{1\over2\ep}\bigg)=\lim_{s\to0}
\bigg({p-\ep\over\ep^{2s} m^2}+{1\over2\ep^{2s+1}}\bigg),
$$
which introduces the series
$$
\ze_2(s,m^2)=\sum_{p=0}^\infty{p\over(p^2+m^2)^s}
\eql{ze2}$$and
$$
\ze_1(s,m^2)=\sum_{p=0}^\infty{1\over(p^2+m^2)^s}
\eql{ze1}$$
so that
$$
\sum_{p=0}^\infty{N_p^{(2)}\over(p^2+m^2)^s}=2\ze_1(s,m^2)
-{1\over m^{2s}}\equiv2\overline\ze_1(s,m^2).
\eql{ze3}$$
Then
$$\sum_{p=0}^\infty N_p^{(2)}
\bigg(\!{p-\ep\over m^2}+{1\over2\ep}\!\bigg)\!=\!
{2\over m^2}\lim_{s\to0}\bigg(\!\ze_2(s,m^2)-\overline\ze_1(s-{1\over2},
m^2)+{m^2\over2}\overline\ze_1(s+{1\over2},m^2)\!\bigg).
\eql{regsum}$$

Although (\peq{ze2}) and (\peq{ze1}) are very standard we
proceed from first principles and write
$$
\ze_2(s,m^2)={1\over\Ga(s)}\int_0^\infty dt\, t^{s-1}\sum_p p
e^{-(p^2+m^2)t}
$$which is related to the \zf\ on a two-sphere. The asymptotic expansion
of the heat-kernel, related to Mulholland's result, is known to be
$$
\sum_{p=0} p e^{-p^2t}\sim {1\over4\pi t}\sum_{n=0,1\ldots}C_nt^n
\eql{as1}$$where
$$
C_0=2\pi,\quad\quad C_n=2\pi{(-1)^{n}\over n!}B_{2n},\quad n>0
$$
in terms of Bernoulli numbers.
Of course the coefficients in (\peq{as1}) are best deduced by applying a
summation formula to the \zf, but it is convenient here to consider them
as given (\eg [\pref{Hurt}]).

The asymptotic expansion of $\ze_2(s,m^2)$ is
then given by (\peq{asympm1}),\mgn{ZETA1.MTH}
$$\eqalign{
\ze_2(s,m^2)&\sim{m^{2-2s}\over2(s-1)}+\sum_{n=1}(m^2)^{1-n-s}
{\Ga(s-1+n)\over2\Ga(s)}{(-1)^nB_{2n}\over n!}\cr
&={m^{2-2s}\over2(s-1)}-{m^{-2s}\over12}+\ldots\cr}
$$
and, similarly,
$$
\overline\ze_1(s,m^2)\sim {\sqrt\pi\over2}{\Ga(s-1/2)\over\Ga(s)}m^{1-2s}.
\eql{as2}$$

In (\peq{regsum}) the first term in brackets is immediately finite,
$$
\ze_2(0,m^2)=-{1\over12}-{m^2\over2},
$$
and the
other two can be combined, asymptotically, to a finite expression using
(\peq{as2}),
$$\eqalign{
\overline\ze_1(s+1/2,m^2)-&{2\over m^2}
\overline\ze_1(s-1/2,m^2)\cr
&\sim{\sqrt\pi\over2}
m^{-2s}\bigg({\Ga(s)\over2\Ga(s+1/2)}-{\Ga(s-1)\over\Ga(s-1/2)}
\bigg)\cr
&={\sqrt\pi\over2}m^{-2s}{s\Ga(s)\over(1-s)\Ga(s+1/2)},\cr}
$$which equals $1/2$ at $s=0$.

We therefore find,
$$
\bigg({d\over dm^2}\bigg)^2\sum_{p=0}N_p^{(2)}\big(\ep-p\log(p+\ep)
\big)\sim-{1\over12m^4}-{1\over4m^2}.
\eql{div}$$and, further,
$$
\bigg({d\over dm^2}\bigg)^2\sum_{p=0}N_p^{(2)}\big(
-{1\over2}\log(\ep)\big)\sim{\pi\over8m^3}.
\eql{div2}$$

Consider now the polynomial part of (\peq{disc2}). The $n=1,2$ and $3$
contributions are
$$\eqalign{
&\bigg({d\over dm^2}\bigg)^2{T_1\over \ep}={25m^2\over32\ep^7}
-{11\over16\ep^5}\cr
&\bigg({d\over dm^2}\bigg)^2{T_2\over \ep^2}=
{15m^4\over4\ep^{10}}-{21m^2\over4\ep^8}
+{13\over8\ep^6}.\cr
&\bigg({d\over dm^2}\bigg)^2{T_3\over \ep^3}=
{12155m^6\over512\ep^{13}}-
{11271m^4\over256\ep^{11}}+{771m^2\over32\ep^9}-{57\over16\ep^7}.\cr}
$$
After multiplying by $N_p^{(2)}$, the summations can be effected using
(\peq{ze3}) and (\peq{as2}).

Going back to (\peq{asympm1}), and writing out the first few terms,
$$
\ze(2,m^2)\sim{1\over4\pi}\bigg({C_0\over m^2}+{\sqrt\pi C_{1/2}\over2m^3}+
{C_1\over m^4}+{3\sqrt\pi C_{3/2}\over 4m^5}+\ldots\bigg),
\eql{fft}$$
elementary arithmetic confirms the heat-kernel expansion derived by
Stewartson and Waechter [\pref{SandW}], and extended recently by Berry
and Howls [\pref{BandH}] to many more terms. We note that Stewartson and
Waechter's method also uses knowledge of just the asymptotic
behaviour of the eigenfunctions, but the information is organised
differently.

The Robin case is similar to the Dirichlet one. The polynomials are,
$$\eqalign{
&\bigg({d\over dm^2}\bigg)^2{R_1\over\ep}={13+12\be\over16\ep^5}
-{35m^2\over32\ep^7}\cr
&\bigg({d\over dm^2}\bigg)^2{R_2\over\ep^2}=-{8\be^2+16\be+15
\over8\ep^6}+3m^2{4\be+9\over\ep^8}-{21m^4\over4\ep^{10}}\cr
&\bigg({d\over dm^2}\bigg)^2{R_3\over\ep^3}={20\be^3+40\be^2+
68\be+63\over16\ep^7}-m^2{140\be^2+532\be+917\over32\ep^9}\cr
&\hspace{************}+m^4{3528\be+14217\over256\ep^{11}}
-m^6{16093\over512\ep^{13}}.\cr}
$$
In two dimensions $\be=h$.

The Robin coefficients, derived from a comparison with (\peq{fft}), agree
with those found by Moss [\pref{Moss}] and so will not be given here.
There
is no difficulty in carrying out the evaluation to any desired order.

Waechter used the method of [\pref{SandW}] to discuss the case $d=3$,
\ie the three-ball. His result was corrected by Kennedy [\pref{Kennedy3}]
who extended the calculation up to $d=6$.

The present method also extends to higher dimensions
without the need for any special summation techniques. Since the
complications are simply arithmetic ones, symbolic manipulation can be used
to advantage.

A method has also recently been developed
by Bordag, Elizalde and Kirsten [\pref{BEK}] that allows any number of
coefficients to be calculated in higher dimensions. Their method is
similar to that of Barvinsky {\it et al} [\pref{Barv,Kam}].

\section {\bf 4. The functional determinant.}
In the general case, Voros [\pref{Voros}] develops the connection between
the two quantities $D(z^2)$ and $\De(z^2)$, where $D$ is given in
(\peq{det}) and the Weierstrassian product is
$$
\De(z^2)=\prod_\la(1-{z^2\over\la})\exp\sum_{k=1}^{[\mu]}{1\over k}
\big({z^2\over\la}\big)^{k},
\eql{weier}$$
which amounts to subtracting the first $[\mu]$ terms of the Taylor
expansion of \break $\ln(1-z^2/\la)$. Formally
$$
\ln\De(z^2)={1\over[\mu]!}\int_0^1d\tau\,(1-\tau)^{[\mu]}\bigg(
{d\over d\tau}\bigg)^{{[\mu]}+1}\sum_\la\ln\big(1-{z^2\tau\over\la}\big).
\eql{remainder}$$
(See also Quine, Heydari and Song [\pref{Quine}].)

Voros' relation is\mgn{define V's}
$$
\ze'(0,0)=
\ln\big({\De(z^2)\over D(z^2)}\big)-\sum_{k=1}^{[\mu]}
FP\ze(k,0){z^{2k}\over k}-\sum_{k=2}^{[\mu]}C_{d/2-k}\big(1+\ldots
+{1\over k+1}\big){z^{2k}\over k!}
\eql{vreln}$$
where $\mu=d/2$ is the order of the sequence of eigenvalues and $FP$
indicates use of the finite part prescription. When
$\mu=1/2$, (\peq{vreln}) is equivalent to (\peq{zedashdiff}).

{}From this equation it is clear that to find $\ze'(0,0)$ it is only
necessary to pick out the $m$--independent part of the right-hand side
which can be done in the infinite $m$ limit.
We note from (\peq{asympm1}) that $\ln D(-m^2)=-\ze'(0,m^2)$ contains no
constant part in this limit and that the two summations in (\peq{vreln})
are also not needed. Hence
$$
\ze'(0,0)=\lim_{m\to\infty}\ln\De(-m^2),
\eql{basicf}$$
and this is our starting formula. Once $\ze'(0,0)$ has been found,
$\ze'(0,m^2)$ can be computed numerically from (\peq{vreln}). A power series
expansion in $m^2$ is easily derived.

\section{\bf 5. Determinant on the disc.}
As an example of the use of this technique consider Dirichlet conditions
on the disc, for which $\mu=1$. In this case (\peq{weier}) reads (the
degeneracy is implied),
$$
\De(-m^2)=
\prod_{p,\al_p}(1+{m^2\over\al_p^2})\exp
\big(-{m^2\over\al_p^2}\big)=\prod_p\big(p!2^pm^{-p}I_p\big)\exp\big(-
{m^2\over4(1+p)}\big),
\eql{2disc}$$
where we have used Rayleigh's explicit results on the sums of inverse
powers of the zeros of Bessel functions rather than the formal subtraction
(\peq{remainder}).

\begin{ignore}
For completeness we derive (\peq{vreln}) for the case in question by
noting that
$$
\bigg({d\over dm^2}\bigg)^2\big(\ze'(0,0)-\ze'(0,m^2)\big)=
\bigg({d\over dm^2}\bigg)^2\ln\De(-m^2)
$$and so, firstly
$$
{d\over dm^2}\big(\ze'(0,0)-\ze'(0,m^2)\big)={d\over dm^2}\ln\De(-m^2)+A
$$where
the constant $A$ can be found from the asymptotic behaviour as $m^2\to
\infty$.
\end{ignore}

{}From (\peq{2disc}) and (\peq{logex1}), asymptotically,
$$\eqalign{
\ln\De(-m^2)\sim\sum_{p=0}^\infty N_p^{(2)}\bigg(p\ln2+\ln p!-\ln\sqrt2\pi
&+\ep-p\ln(p+\ep)\cr
&-\ln\sqrt\ep+\sum_n{T_n(t)\over\ep^n}-{m^2\over4(1+p)}\bigg)\cr}
\eql{logdet1}$$
We would like to do the sum over $p$ and take the large $m^2$ limit.
The integral representation of $\log p!$ is employed to give
$$\eqalign{
\ln\De(-m^2)\sim&-\ln\sqrt(2\pi)+m-{1\over2}\ln m +\sum_{p=0,n=2}N_p^{(2)}
{T_n(t)\over \ep^n}\cr
&-{1\over12}\sum_{p=0}N_p^{(2)}\big({1\over\ep}-{5m^2\over2\ep^3}\big)
+2\sum_{p=1}^\infty\bigg[p\ln{2p\over p+\ep}+\ep-p\cr
&\hspace{******}-{m^2\over4p}-{1\over2}\ln{\ep\over p}
+\int_0^\infty\bigg({1\over2}
-{1\over t}+{1\over e^t-1}\bigg){e^{-tp}\over t}\,dt\bigg]\cr}
\eql{logdet2}$$
where some initial terms have been extracted for convenience of doing the
summation and taking the asymptotic limit.
It is readily confirmed by expanding in $m/p$ that the sum over $p$
converges.

Consider firstly,
$$\lim_{m\to\infty}
{5m^2\over24}\sum_{p=0}{N_p^{(2)}\over\ep^3}
$$
which can be evaluated using (\peq{as2}) to give $5/12$.

Next look at
$$
\sum_{p=1}\bigg(\int_0^\infty\bigg({1\over2}-{1\over t}
+{1\over e^t-1}\bigg){e^{-tp}\over t}\,dt-{1\over12\ep}.
$$
The sum over the integral diverges because of the behaviour at $t=0$.
Adding and subtracting the offending term, and performing the sum, yields

$$
\int_0^\infty\bigg({1\over2}-{1\over t}-{t\over12}
+{1\over e^t-1}\bigg){1\over t(e^t-1)}\,dt+{1\over12}\sum_{p=1}
\big({1\over p}-{1\over\ep}\big).
$$

Dealing with the final summation first, we make use of the standard
formula [\pref{Watson2}]\mgn{INFSUM5}
$$\eqalign{
\sum_{p=1}^\infty
\big({1\over p}-{1\over\ep}\big)&=
\log m+{1\over2m}-2\sum_{p=1}^\infty K_0(2\pi mp)+\ga-\log2\cr}
\eql{bessum1}$$
the last two terms of which are the required constant contribution.
The integral term can be regularised as
$$
\lim_{s\to0}\int_0^\infty\bigg({1\over2}-{1\over t}-{t\over12}
+{1\over e^t-1}\bigg){t^{s-1}\over e^t-1}\,dt
=2\ze_R'(-1)+{1\over2}\ln\sqrt{2\pi}-{1+\ga\over12}.
\eql{regint}$$

The remaining terms in the summation in (\peq{logdet2}) are now evaluated
in piecemeal fashion.

The $\ln(\ep/p)$ term in (\peq{logdet2}) can be found
by differentiating the particular Watson-Kober formula
[\pref{Kober,Watson2}]
$$\eqalign{
\ze_R(2s)=\sum_{p=1}^\infty\big({1\over p^{2s}}-&{1\over\ep^{2s}}\big)
+{2\pi^sm^{(1-2s)/2}\over\Ga(s)}\sum_{n=1}^\infty n^{(2s-1)/2}
K_{(2s-1)/2}(2\pi nm)\cr
&-{m^{-2s}\over2}+{m^{1-2s}\Ga\big(s-1/2\big)\sqrt\pi\over2\Ga(s)}\cr}
\eql{wk}$$
valid for $\Real s>-1/2$, in particular near $s=0$. We find
$$
\sum_{p=1}^\infty\big(\ln\ep-\ln p\big)={\pi m\over2}-\ln\sqrt(2\pi)-
{1\over2}\ln m+{1\over2}\ln\big(1-e^{-2\pi m}\big).
\eql{logsum}$$

Alternatively, the asymptotic part only of this follows by differentiating
$$
\sum_{p=1}^\infty\big({1\over p^{2s}}-{1\over\ep^{2s}}\big)=
\ze_R(2s)-\overline\ze_1(s)+{1\over2m^{2s}}
$$
and using (\peq{as2}). The same goes for (\peq{bessum1}).

To treat the $(\ep-p)$ term, twice the following term, \ie $m^2/2p^2$,
could be replaced by $m^2/2\ep^2$, plus a known correction, using
(\peq{bessum1}), so allowing the earlier result (\peq{regsum}) to be
applied. However it is preferred here to use directly Watson's
general formula,
$$\eqalign{
\sum_{p=1}^\infty\bigg({1\over(p^2+m^2)^s}&-\sum_{n=0}^{M-1}\comb{-s}n
{m^{2n}\over p^{2s+2n}}\bigg)={\sqrt\pi\Ga\big(s-1/2\big)
m^{1-2s}\over2\Ga(s)}-{1\over2}m^{-2s}\cr
&+{2\pi^s\over\Ga(s)}m^{(1-2s)/2}\sum_{p=1}
^\infty p^{(2s-1)/2}K_{(2s-1)/2}(2\pi mp)\cr
&-\sum_{n=0}^{M-1}\comb{-s}n m^{2n}\ze_R(2s+2n)\cr}
\eql{watson3}$$
which is valid for $\Real s>1/2-M$ and plays an important role in the
sequel. If $M$ is set equal to 1,
one gets (\peq{wk}) and the $s\to1/2$ limit then yields (\peq{bessum1}).
We now need to take $M=2$. This gives
$$\eqalign{
\sum_{p=1}^\infty\bigg({1\over\ep^{2s}}&-{1\over p^{2s}}+s{m^2\over
p^{2s+2}}\bigg)={\sqrt\pi\Ga\big(s-1/2\big)
m^{1-2s}\over2\Ga(s)}-{1\over2}m^{-2s}\cr
&+{2\pi^s\over\Ga(s)}m^{(1-2s)/2}\sum_{p=1}
^\infty p^{(2s-1)/2}K_{(2s-1)/2}(2\pi mp)\cr
&\hspace{*****}-\ze_R(2s)+sm^2\ze_R(2s+2)\cr}
\eql{watson4}$$
valid for $\Real s>-3/2$. In particular at $s=-1/2$
$$
\sum_{p=1}^\infty\big(\ep-p-{m^2\over 2p}\big)
={m^2\over4}(2\ln2+1)-{1\over2}\ga m^2-{1\over2}m-{1\over4}m^2\ln m^2
-{m\over\pi}\sum_{p=1}^\infty {K_1(2\pi mp)\over p}+{1\over12}
\eql{watson5}$$
which, in conjunction with (\peq{bessum1}), provides an
alternative derivation of (\peq{regsum}) and (\peq{div}).

Incidentally, the integral representation
$$
\Ga(1/2-\nu)K_\nu(z)=\sqrt\pi(z/2)^{-\nu}\int_1^\infty e^{-zt}(t^2-1)^
{-\nu-1/2}\,dt
\eql{rep1}$$
allows the summation to be performed,
$$\eqalign{
-{m\over\pi}\sum_{p=1}^\infty {K_1(2\pi mp)\over p}&=
-2m^2\int_1^\infty{1\over e^{2\pi mt}-1}(t^2-1)^{1/2}\,dt\cr
&={m\over\pi}
\int_1^\infty\ln\big(1-e^{-2\pi mt}\big){t\over(t^2-1)^{1/2}}\,dt.\cr}
\eql{k1sum}$$
Equation (\peq{watson5}) then agrees with a formula derived by Elizalde,
Leseduarte and Zerbini [\pref{ELZ}] in a different connection by a longer
method.

Harking back to (\peq{logdet2}), all terms have now been
dealt with except for
$$
\sum_{p=1}^\infty\bigg(p\ln{2p\over p+\ep}+{m^2\over4p}\bigg).
$$

If this expression is differentiated, (\peq{bessum1}) and (\peq{watson5})
employed and the result integrated we find,\mgn{INFSUM1, INFSUM6.TRU}
$$\eqalign{
\sum_{p=1}^\infty&\bigg(p\ln{2p\over p+\ep}+{m^2\over4p}\bigg)=
{1\over8}\big(m^2+{1\over3}\big)\ln m^2+{1\over4}m^2\big(\ga-\ln2\big)\cr
&+{m\over\pi}\sum_p {K_1(2\pi mp)\over p}
+{1\over2\pi^2}\sum_p{K_0(2\pi mp)\over p^2}-{1\over12}\ln2-\ze_R'(-1).\cr}
\eql{fsum}$$
where the constant of integration has been found by setting $m$ to zero.
This is allowed because (\peq{fsum}) is exact. From (\peq{watson5})
$$
\lim_{m\to0} {m\over\pi}\sum_{p=1}^\infty {K_1(2\pi mp)\over p}={1\over12}
\eql{k1sum2}$$
which can be derived from (\peq{k1sum}) and also by replacing $K_1(z)$
by its leading term, $\sim 1/z$, as $z\to0$.

Replacing $K_0(z)$ by its leading term as $z\to0$, \ie $-\ln(z/2)-\ga$,
leads to
$$\eqalign{
\lim_{m\to0}{1\over2\pi^2}\sum_{p=1}^\infty{K_0(2\pi mp)\over p^2}&=
-{1\over12}
\big(\ln(\pi m) +\ga\big)+{1\over2\pi^2}\ze_R'(2)\cr
&=\ze_R'(-1)+{1\over12}\big(\ln2-1-\ln m\big)\cr}
\eql{k0sum2}$$
where we have used
$$
{1\over2\pi^2}\ze_R'(2)=\ze_R'(-1)+{1\over12}(\ln 2\pi+\ga-1).
$$
\mgn{$\ze'(2)=-0.93754825,\break \ze'(-1) = -0.16542114$}
These results enable equation (\peq{fsum}) to be completed.

As an aside, a useful expression arises from the Fourier
transform representation
$$
K_\nu(az)=\pi^{-1/2}(2z/a)^\nu\Ga(\nu+1/2)\int_0^\infty{\cos at\over
(t^2+z^2)^{\nu+1/2}}\,dt.
\eql{rep2}$$
Thus
$$
\sum_p{K_0(2\pi mp)\over p^2}=\int_0^\infty\sum_1^\infty{\cos pt\over p^2}
{dt\over(t^2+4\pi^2 m^2)^{1/2}}.
$$

The sum over $p$ gives a Bernoulli polynomial in the range $0\le t\le2
\pi$ and
splitting up the integral into $2\pi$ lengths we find, after rescaling,
$$
{1\over\pi^2}\sum_{p=1}^\infty{K_0(2\pi mp)\over p^2}
=\int_{-1}^1B_2(t)\sum_{n=1,3,}^\infty{dt\over
\big((t+n)^2+4m^2)^{1/2}}
\eql{int2}$$
where
$$
B_2(t)={3t^2-1\over12}.
$$

The integrals can be performed, leading to the summation
\begin{ignore}
$$
{1\over\pi^2}\sum_p{K_0(2\pi mp)\over p^2}
=m-{1\over6}\ln m-2\sum_{n=1}^\infty
\bigg(n\ln\big((n^2+m^2)^{1/2}+n\big)
-(n^2+m^2)^{1/2}\bigg)
\eql{k0sum3}$$
\end{ignore}

$$\eqalign{
{1\over\pi^2}&\sum_{p=1}^\infty{K_0(2\pi mp)\over
p^2}={1\over24}\sum_{n=1,3,}^\infty
\bigg[2\big(3n^2-1\big)\ln\bigg({\big((n+1)^2+4m^2\big)^{1/2}+n+1
\over\big((n-1)^2+4m^2\big)^{1/2}+n-1}\bigg)\cr
&-3\big((3n-1)\sqrt((n+1)^2+4m^2)-(3n+1)\sqrt((n-1)^2+4m^2)\big)\bigg],\cr}
\eql{k0sum4}$$
which is convenient numerically as a means of checking the foregoing
expansions.

These results can be generalised to sums of the form
$\sum_pK_0(zp)/p^{2n}$. We give only the formula
$$\eqalign{
\lim_{m\to0}&\sum_{p=1}^\infty {K_0(2\pi mp)\over p^{2n}}\cr
&={(-1)^{n+1}2^{2n-1}\pi^{2n}\over(2n)!}\bigg(B_{2n}
\big(\ln2-\ln m+\ga-\psi(2n)\big)+2n\ze_R'(1-2n)\bigg)\cr}
\eql{k0sum5}$$
and also note that\mgn{check defs of $B_n$}
$$
\lim_{m\to0}{m^l\over\pi^n}\sum_{p=1}^\infty {K_l(2\pi mp)\over p^n}
=2^{n+l-2}{(l-1)!\over(n+l)!}B_{n+l}
\eql{k0sum6}$$
when $l+n$ is even.

Incidentally, taking the $m\to0$ limit of (\peq{k0sum4}) and comparing
with (\peq{k0sum2}) gives an expression for $\ze'_R(-1)$\mgn{ZEDASH.TRU}
$$\eqalign{
\ze'_R(-1)&={1\over24}\sum_{n=3,5,}^\infty\bigg((3n^2-1)\ln\big({n+1
\over n-1}\big)-6n\bigg)-{1\over6}\cr
&={1\over6}\sum_{k=3,5,}^\infty{k-1\over k(k+2)}\big((1-2^{-k})
\ze_R(k)-1\big)-{1\over6}.\cr}
\eql{zedsum}$$
Numerically, from (\peq{zedsum}) we find
$\ze_R'(-1)\approx-0.165421143700045092139$.
A summation for $\ze_R'(-1)$ is given by Elizalde and Romeo [\pref{EandR}].
\mgn{BESSUM2.MTH}Similar summations can be found for $\ze'_R(1-2n)$.

All parts of (\peq{logdet2}) have now been evaluated. Collecting the
constant terms we arrive at
$$
\ze'_{\rm2ball}(0,0)=2\ze_R'(-1)+{1\over6}\ln2+{5\over12}
-{1\over2}\ln\pi\approx\,0.773714,
\eql{zedashd}$$
in agreement with Weisberger [\pref{Weisbergerb}].

As a tactical point we note that the equations we have developed contain
far more information than needed for the specific purpose of picking out
the $m$--independent part and so, from this point of view, the method is
not very economical. A more streamlined version is presented in section 8
where only the essentials are retained.
\section{\bf 6. The Robin case.}
In this case the appropriate expansion is (\peq{logex3}) and we firstly
need $\sum 1/\al^2$.
It is straightforward to derive,
$$\eqalign{
\sum_{\al_p}{1\over\al_p^2}&={1\over4(p+1)}\bigg(1+{2\over\be+p}\bigg)\cr
&={1\over8},\quad \be=0,\,p=0.\cr}
\eql{robroot}$$

We remark that Lamb's formula, [\pref{Lamb}]
$$
\sum_\al{1\over\al^2+\be^2-p^2}={1\over2(\be+p)},\quad\be>0,
$$
is not enough to prove results like (\peq{robroot}) but does provide a
test when $\be=p$. A rough, numerical check of (\peq{robroot}), and
similar
formulae, can be obtained in the $p=0$ case by using the roots given in
Table III of Carslaw and Jaeger [\pref{CandJ}].

Instead of (\peq{2disc}) we have
$$
\De(-m^2)=\prod_{p=0}^\infty\big({p!2^pm^{-p}\over\be+p}F_p\big)\exp\bigg(
-{m^2\over4(1+p)}\big(1+{2\over\be+p}\big)\bigg)
$$
and so
$$\eqalign{
\ln\De(-m^2)\sim\sum_{p=0}^\infty\bigg(p\ln2+\ln p!&-\ln\sqrt(2\pi)
-\ln(\be+p)+\ep-p\ln(p+\ep)\cr
&-\ln\sqrt\ep+\sum_n{S_n\over p^n}-{m^2\over4(1+p)}
\big(1+{2\over\be+p}\big)\bigg).\cr}
$$

Corresponding to (\peq{logdet2}) we find ($\be\ne0$)
$$\eqalign{
\ln\De(-m^2)\sim&-\ln\sqrt(2\pi)-\ln\beta+m-{1\over2}\ln m
+\sum_{p=0,n=2}N_p^{(2)}{R_n(t)\over \ep^n}\cr
&+{1\over12}\sum_{p=0}N_p^{(2)}\big({12\be-1\over\ep}
-{7m^2\over2\ep^3}\big)
+2\sum_{p=1}^\infty\bigg[p\ln{2p\over p+\ep}+\ep-p\cr
&\hspace{*******}-{m^2\over4p}\big(1+{2\over\be+p-1}\big)
+{1\over2}\ln{\ep\over p}-\ln(1+\be/p)\cr
&\hspace{*****************}+\int_0^\infty\bigg({1\over2}-{1\over t}
+{1\over e^t-1}\bigg){e^{-tp}\over t}\,dt\bigg]\cr}
\eql{logdetr}$$
which can be reorganised in the same way that (\peq{logdet2}) was.

The only sum not covered by previously given expressions is
$$
\sum_{p=1}^\infty\bigg({\be\over\ep}-\ln\big(1+{\be\over p}\big)\bigg)
$$
which, using (\peq{bessum1}), can be converted to
$$
\sum_{p=1}^\infty\bigg({\be\over p}-\ln\big(1+{\be\over p}\big)\bigg)
=\ln\Ga(1+\be)+\ga\be.
\eql{lsum}$$
Putting all the pieces together and selecting the $m$-independent part
yields, if $\be\ne0$,
$$
\ze'_{Ro}(0,0)=2\ze'_R(-1)-{7\over12}-{5\over6}\ln2-{1\over2}\ln\pi
+2\ln\Ga(1+\be)+2\be\ln2-\ln\beta.
\eql{zedashr}$$

The case when $\be$ is zero must be treated separately because of the
appearance of extra zero modes. The $p=0$ term contributes an additional
$\ln2$, which
is clear from the fact that $J_0(z)=-zJ_1(z)\sim z^2/2$ and the result is
the Neumann determinant,
$$
\ze'_{N}(0,0)=2\ze'_R(-1)-{7\over12}+{1\over6}\ln2-{1\over2}\ln\pi.
\eql{zedashn}$$
\section{\bf 7. The four-ball.}

We will pursue the calculation of the functional determinant in detail
only for $\rB^4$. In this case the order $\mu$ is two and equation
(\peq{weier}) reads
$$\eqalign{
\De(-m^2)&=\prod_{p,\al_p}(1+{m^2\over\al_p^2})\exp
\big(-{m^2\over\al_p^2}+{m^4\over2\al_p^4}\big)\cr
&=\prod_p\big(p!2^pm^{-p}I_p\big)\exp\big(-{m^2\over4(1+p)}+{m^4
\over32(1+p)^2(2+p)}\big)\cr}
\eql{4sphere}$$
and (\peq{logdet1}) is
$$\eqalign{
\ln\De(-m^2)\sim\sum_{p=1}^\infty p^2\bigg(p\ln2&+\ln p!-\ln\sqrt2\pi
+\ep-p\ln(p+\ep)\cr
&-\ln\sqrt\ep+\sum_n{T_n(t)\over \ep^n}
-{m^2\over4(1+p)}+{m^4\over32(1+p)^2(2+p)}\bigg).\cr}
\eql{logdet4}$$

The situation now is that more terms in the summation over the polynomials
$T_n$ become relevant.

(\peq{logdet2}) is\mgn{ZETA1.MTH, OLVERUS.MTH, ESS4.MTH, SUMD.MTH}
$$\eqalign{
\ln\De(-m^2)\sim&\sum_{p=1,n=4}p^2{T_n(t)\over \ep^n}+\sum_{p=1}
\bigg({5m^2\over24\ep}-{5m^4\over24\ep^3}+{9m^4\over16\ep^4}
-{5m^6\over16\ep^6}-{m^2\over4\ep^2}\cr
&+{1\over360}\bigg({1\over\ep}-{1\over p}\bigg)+{181m^2\over720\ep^3}
-{1349m^4\over960\ep^5}+{2431
m^6\over1152\ep^7}-{1105m^8\over1152\ep^9}\bigg)\cr
&+\sum_{p=1}^\infty p^2\bigg[p\ln{2p\over p+\ep}+\ep-p
-{m^2\over4(1+p)}+{m^4\over32(1+p)^2(2+p)}\cr
&-{1\over2}\ln{\ep\over p}+{1\over12}\big({1\over p}-{1\over\ep}\big)
\cr
&+\int_0^\infty\bigg({1\over2}-{1\over t}-{t\over12}+{t^3\over720}
+{1\over e^t-1}\bigg){e^{-tp}\over t}\,dt\bigg].\cr}
\eql{logdet5}$$

Rearranging the summations of the Rayleigh terms for future use,
$$\eqalign{
\sum_{p=1}^\infty&p^2\bigg(-{m^2\over4(1+p)}+{m^4\over32(1+p)^2(2+p)}
\bigg)\cr
&={1\over4}\sum_{p=1}^\infty \bigg[m^2\bigg(1-p-{1\over p}
+{1\over p(1+p)}\bigg)+{1\over8}m^4
\bigg({1\over p}-{3\over p(1+p)}+{1\over p^2(1+p)}\bigg)\bigg]\cr
&={1\over4}\sum_{p=1}^\infty \bigg[m^2\bigg(1-p-{1\over p}
\bigg)+{1\over8p}m^4\bigg]+{m^2\over4}
+m^4\bigg({\pi^2\over6}-4\bigg)\cr}
\eql{Rayl}$$

The convergent summations would be needed for a finite mass calculation but
are irrelevant here and (\peq{Rayl}) can thus be replaced by
$$
-{1\over4}\sum_{p=1}^\infty\bigg[ m^2\bigg(p-1+{1\over p}\bigg)-{1\over8}
{m^4\over p}\bigg]+O(m^2).
$$

The formulae needed to analyse (\peq{logdet5}] are extensions of those used
earlier. We set $M$ equal to 3 in (\peq{watson3}), to get
$$\eqalign{
\sum_{p=1}^\infty\bigg({1\over\ep^{2s}}&-{1\over p^{2s}}+s{m^2\over
p^{2s+2}}-{s(s+1)\over2}{m^4\over p^{2s+4}}\bigg)
={\sqrt\pi\Ga\big(s-1/2\big)m^{1-2s}\over2\Ga(s)}-{1\over2}m^{-2s}\cr
&+{2\pi^s\over\Ga(s)}m^{(1-2s)/2}\sum_{p=1}
^\infty p^{(2s-1)/2}K_{(2s-1)/2}(2\pi mp)\cr
&\hspace{*****}-\ze_R(2s)+sm^2\ze_R(2s+2)-{s(s+1)\over2}m^4\ze_R(2s+4)\cr}
\eql{watson6}$$ and then let $s$ tend to $-3/2$, yielding \mgn{INFSUM10.TRU}
$$\eqalign{
\sum_{p=1}^\infty\big(\ep^3-p^3-{3m^2\over 2}p-{3\over8}{m^4\over p}\big)
=&-\ze_R(-3)+{1\over8}m^2-{1\over2}m^3-{3\over16}m^4\ln m^2-{3\over8}m^4
\ga\cr
&+{3\over32}m^4(4\ln2+3)+{3m^2\over2\pi^2}\sum_{p=1}^\infty
{K_2(2\pi mp)\over p^2}.\cr}
\eql{watson7}$$

Another necessary equation follows upon differentiating (\peq{watson4})
with respect to $s$ and then setting $s=-1$,\mgn{LOGSUM1.MTH}
$$\eqalign{
\sum_p\big(2\ep^2(\ln p-\ln\ep)+m^2\big)=&m^2\ln m^2-
{1\over2}m^2-2m^2\ln\sqrt(2\pi)-{2\pi\over3}m^3\cr
&-{m\over\pi}\sum_p{e^{-2\pi mp}\over p^2}
-{1\over2\pi^2}\sum_p{e^{-2\pi mp}\over p^3}-2\ze_R'(-2).\cr}
\eql{watson8}$$

As a small check, we note that $\ze'_R(-2)=-\ze_R(3)/4\pi^2$ and the
right hand side correctly vanishes when $m$ is zero.

The final equation we need is
$$\eqalign{
\sum_{p=1}^\infty\bigg(&p^3\ln\big({2p\over p+\ep}\big)+{m^2 p\over4}
-{3m^4\over32p}\bigg)={m^4\over32}(3\ga-3\ln2+1)+{1\over12}m^3+{7\over48}
m^2\cr
&-{1\over2}\ze_R(-3)\ln m^2
-{11\over64}m^4\ln m^2-{m^3\over\pi}\sum_{p=1}^
\infty{K_3(2m\pi p)\over p}\cr
&\hspace{**}-{3m\over4\pi^3}\sum_{p=1}^\infty{K_1(2m\pi p)\over p^3}
-{3\over4\pi^4}\sum_{p=1}^\infty{K_0(2m\pi p)\over p^4}+
{1\over120}\ln2-\ze_R'(-3),\cr}
\eql{watson9}$$
which is obtained in a similar way to (\peq{fsum}) and involves using
(\peq{k0sum5}) and (\peq{k0sum6}).

The integral is again regularised
$$\eqalign{
&\lim_{s\to0}\int_0^\infty\bigg({1\over2}-{1\over t}-{t\over12}+{t^3
\over720}
+{1\over e^t-1}\bigg)t^{s-1}{d^2\over dt^2}{1\over e^t-1}\,dt .\cr}
\eql{regint2}$$

The power terms are most easily dealt with by partial integration which
gives
$$
\int_0^\infty t^\mu t^{s-1}{d^j\over dt^j}{1\over e^t-1}\,dt=
(-1)^j\Ga(\mu+s)\ze_R(\mu-j+s).
\eql{powers1}$$

For the specific case here, $j=2$ and the finite part of the power terms
is
$$
\ze_R'(-3)+{1\over2}\ze_R'(-2)-{1\over120}\ga+{7\over360}.
\eql{powers2}$$

The remaining piece of (\peq{regint2}) is
$$\eqalign{
\int_0^\infty& {1\over e^t-1}t^{s-1}{d^2\over dt^2}{1\over e^t-1}\,dt\cr
&=\int_0^\infty t^{s-1}\bigg({2\over(e^t-1)^4}+{3\over(e^t-1)^3}
+{1\over(e^t-1)^2}\bigg)\,dt.\cr}
\eql{bar}$$

These integrals are examples of a particular Barnes zeta function,

$$\eqalign{
\ze_r(s,a)&={i\Ga(1-s)\over2\pi}\int_L{e^{z(r-a)}(-z)^{s-1}\over
(e^z-1)^r}\,dz\cr
&=\sum_{n=0}^\infty\comb{n+r-1}{r-1}{1\over{(a+n)}^s}\,,\quad \Real s>r.
\cr}
\eql{hemibarnes}$$ In the present case $a=r$ and the lower limit can be
adjusted so that
$$
\ze_r(s,r)=\sum_{n=1}^\infty\comb {n-1}{r-1}{1\over n^s}.
$$

Expanding the binomial
coefficient as a polynomial in $n$ using Stirling numbers
$$
\comb{n-1}{r-1}={1\over(r-1)!}\sum_{k=1}^r S^{(k)}_r\,n^{k-1}
$$gives
$$
\ze_r(s,r)={1\over (r-1)!}\sum_{k=1}^rS^{(k)}_r\,\ze_R(s-k+1)
\eql{barnexp}$$
in terms of the Riemann \zf.

Explicitly for $d=4$ we need
$$\eqalign{
&\ze_2(s,2)=\ze_R(s-1)-\ze_R(s)\cr
&\ze_3(s,3)={1\over2}\ze_R(s-2)-{3\over2}\ze_R(s-1)+\ze_R(s)\cr
&\ze_4(s,4)={1\over6}\ze_R(s-3)-\ze_R(s-2)+{11\over6}\ze_R(s-1)-\ze_R(s).
\cr}
$$

Of course, one could legitimately
leave the answer in terms of the Barnes\break \zf\ and the multiple
$\Ga$--function.

The finite part of (\peq{bar}) is evaluated at $s=0$ to give\mgn{CHECK}
$$\eqalign{
2\ze'_4(0,4)+3\ze'_3(0,3)+\ze'_2(0,2)-&\ga(2\ze_4(0,4)+3\ze_3(0,3)
+\ze_2(0,2))=\cr
&{1\over3}\ze_R'(-3)-{1\over2}\ze_R'(-2)+{1\over6}\ze'_R(-1)+{1\over90}
\ga\cr}
\eql{fp1}$$and this has to be added to (\peq{powers2}) to get the total
contribution from the integral term.

With these and the earlier equations it is possible to evaluate the
summations
over $p$ in (\peq{logdet5}) and to extract the $m$--independent part.
A point to note is that the convergent terms coming from the $T_3$
polynomial all contribute to this constant term using (\peq{as2}).

Regarding this, a minor calculational point should be remarked. Because of
the $p^2$ factor, the sum over $p$ can be extended to $p=0$ when
convenient.
This means that a factor of $N_p^{(2)}/2$ can be inserted without penalty so
allowing the simple asymptotic form (\peq{as2}) to be used. A more refined
approach is given in section 8.

Combining all $m$-independent contributions by hand we obtain
$$
\ze'_{\rm 4ball}(0)={1\over3}\ze_R'(-3)-{1\over2}\ze'(-2)
+{1\over6}\ze_R'(-1)+{1\over90}\ln2+{173\over30240}\approx\,0.002869,
$$
agreeing with a previous result [\pref{DandA}].

We note that there is cancellation of the $\ze_R'$ terms in
(\peq{powers2})
with those from the summations, leaving just those in (\peq{fp1}).
The $\ga$ terms always cancel. These facts are generic as will now be shown
\section{\bf 8. The d-Ball.}
With the experience gained in the previous sections, the calculation
can be readily extended to arbitrary dimensions.

The degeneracy is expanded as usual, (for $d>2$)
$$
N_p^{(d)}=\sum_{j=1}^{d-2}H_j^{(d)} p^j,
\eql{degend}$$
which is an odd polynomial for $d$ odd and even for $d$ even. The
coefficients are Stirling numbers.

For even balls, it suffices to discuss
$$\eqalign{
\ln\De(-m^2)&\sim\sum_{p=1}^\infty p^{2\nu}\bigg(p\ln2+\ln p!-\ln\sqrt2\pi
+\ep-p\ln(p+\ep)\cr
&\hspace{********}-\ln\sqrt\ep+\sum_{n=1}^\infty{T_n(t)\over \ep^n}
-{\rm Ray}(m,p)\bigg).\cr
&=\sum_{p=1}^\infty p^{2\nu}\bigg[p\ln{2p\over p+\ep}+\ep-p
-{1\over2}\ln{\ep\over p}+\sum_{n=1}^\infty{T_n(t)\over \ep^n}\cr
&\hspace{***}-{\rm Ray}(m,p)+\int_0^\infty\bigg({1\over2}
-{1\over t}+{1\over e^t-1}\bigg){e^{-tp}\over t}\,dt\bigg],\cr}
\eql{logdet7}$$ where ${\rm Ray}(m,p)$ are the `Rayleigh' terms.

It is enough to decide what terms are {\it irrelevant} for the
$m$-independent part of the large $m$ limit and we first note that the
Rayleigh terms are so irrelevant. Thus it is not necessary to find the
sums of inverse powers of the eigenvalues, interesting though these
may be.

Again we go through the various contributions. The analysis is somewhat
involved and fragmented, but not without interest. Much of the difficulty
is bookeeping.

First consider
$$\eqalign{
p^{2\nu}(&\ep^\al-p^\al)=\big(\ep^2-m^2\big)^\nu(\ep^\al-p^\al)\cr
&=\sum_{q=0}^\nu\comb \nu q (-1)^{\nu-q}m^{2\nu-2q}
\big(\ep^{2q+\al}-p^{2q+\al}\big)
+\sum_{q=0}^{\nu-1}\comb \nu q (-1)^{\nu-q}m^{2\nu-2q}p^{2q+\al}\cr}
\eql{epexp}$$
for $\al=1$.

{}From (\peq{watson3}) the relevant part of
$\sum_p\big(\ep^{2q+1}-p^{2q+1}\big)$
is just $-\ze_R(-2q-1)$ and so that of $\sum_pp^{2\nu}(\ep-p)$ equals
$-\ze_R(-2\nu-1)$ since terms involving $m$ can be ignored and the second
sum in (\peq{epexp}) is ultimately irrelevant.

The next contribution is
$$
-{1\over2}\sum_pp^{2\nu}\ln{\ep\over p}
$$
whose relevant part, $-\ze_R'(-2\nu)/2$, is obtained by differentiating
(\peq{watson3}) and setting $s=-\nu$, after using $\ep^2=p^2+m^2$.

Now consider
$$
\sum_pp^{2\nu+1}\ln{2p\over p+\ep}
\eql{hsum}$$
which is the hardest sum to evaluate. Our method involves differentiation
with respect to $m^2$, use of the Watson-Kober formula (\peq{watson3}) and
then a back integration, the relevant term emerging as a constant of
integration.

It is clear from the examples in previous sections that only
the Bessel functions in (\peq{watson3}) ultimately contribute
to the relevant term. However we write out (\peq{watson3}) in general
form, setting $M=q+2$ and $s=-(2q+1)/2$ ($q\in\oZ$) to give
$$\eqalign{
\sum_{p=1}^\infty&\bigg[\big(\ep^{2q+1}-p^{2q+1}\big)-\sum_{l=1}^{q+1}
\comb{q+1/2}lm^{2l}p^{2l+1+2q}\bigg]=-\ze_R(-2q-1)\cr
&-{1\over2}m^{2q+1}-{\Ga(q+1/2)\over2\sqrt\pi}{(1+2q)\over(q+1)!}m^{2q+2}
\bigg(\ga-\ln2+\sum_{k=1}^{1+q}\big({1\over2k+1}-{1\over k}\big)\bigg)\cr
&-\sum_{l=1}^q\comb{q+1/2}lm^{2l}\ze_R(2l-2q-1)+{2\pi^{-q-1/2}m^{1+q}
\over\Ga(-q-1/2)}\sum_{p=1}^\infty{K_{1+q}(2\pi mp)\over p^{1+q}},\cr}
\eql{watson10}$$
only the last term of which is actually needed.

Returning to the required contribution, (\peq{hsum}), sufficient members
($\nu$ in number) of the Taylor expansion are subtracted to render the
summation finite. We denote this regularisation by $\sumstar{}$. Then we
find, using (\peq{epexp}),
$$\eqalign{
{d\over dm^2}\sumstart{p}{}p^{2\nu+1}\ln{2p\over p+\ep}&=
{1\over2}\sum_{q=0}^{\nu-1}(-1)^{\nu-q}\comb {\nu+1}{q+1}
m^{2\nu-2q-2}\sumstart{p}{}\big(\ep^{2q+1}-p^{2q+1}\big)\cr
&\hspace{***}+{1\over2m^2}\sumstart{p}{}\big(\ep^{2\nu+1}-p^{2\nu+1}\big)-
(-1)^\nu{m^{2\nu}\over2}\sumstart{p}{}\big({1\over\ep}-{1\over p}\big),\cr}
\eql{diffregsum}$$
where $\sumstar{} \big(\ep^{2q+1}-p^{2q+1}\big)$ is just
the left-hand side of equation (\peq{watson10}).

Integrating (\peq{diffregsum}) we see that all terms in (\peq{watson10}),
apart possibly from the Bessel function ones, will produce irrelevant
contributions and so the problem is reduced to evaluating the massless
limit of the summations
$$
\sum_{p=1}^\infty\int m^{2\nu-q}{K_{1+q}(2\pi mp)\over p^{1+q}}\,dm
\eql{mint}$$
where $q$ ranges from $-1$ to $\nu$.

The integrations can be performed using a recursion given by Watson
[\pref{Watson1}] which reduces the $2\nu$ exponent to zero in steps of 2
finally allowing use of the integral
$$
\int x^{-q}K_{1+q}(x)\,dx=-x^{-q}K_q(x).
$$

The result of the recursion is given in terms of terminating hyperbolic
Lommel functions\mgn{SIGNS} [\pref{Erdelyi}], {\bf7.14.1} (7),
$$
\int x^{2\nu-q} K_{1+q}(x)=-2\nu xS_{2\nu-q-1,q}(x)K_{1+q}(x)-
xS_{2\nu-q,q+1}(x)K_q(x).
\eql{intq}$$

Written out, the Lommel functions are
\begin{ignore}
$$\eqalign{
S_{2\nu-q-1,q}(x)=x^{2\nu-q-2}\,\bigg(1-&{\big[(2\nu-q-2)^2-q^2\big]
\over x^2}\cr
&+{\big[(2\nu-q-2)^2-q^2\big]\big[(2\nu-q-4)^2-q^2\big]\over x^4}
+\ldots\bigg)\cr}
$$
\end{ignore}
$$\eqalign{
S_{2\nu-q-1,q}(x)=x^{2\nu-q-2}\,\bigg(1+&{4(\nu-1)(\nu-q-1)\over x^2}\cr
&+{4^2(\nu-1)(\nu-2)(\nu-q-1)(\nu-q-2)\over x^4}+\ldots\bigg)\cr}
$$
and\mgn{LOMMEL.MTH}
$$\eqalign{
S_{2\nu-q,q+1}(x)=x^{2\nu-q-1}\,\bigg(1+&{4\nu(\nu-q-1)\over x^2}\cr
&+{4^2\nu(\nu-1)(\nu-q-1)(\nu-q-2)\over x^4}+\ldots\bigg).\cr}
$$

Since the massless limit has to be taken, some simplification can be
made at this point. The leading term in $K_q(x)$, $q>0$, as $x\to0$ is
$2^{q-1}(q-1)!/x^q$, while $K_0(x)\sim -(\ln(x/2)+\ga)$. For $q=-1$ and
$q=0$ it is easily checked that the
$K_0$ terms in (\peq{intq}) go out as $x\to0$ and that, for
$0\le q\le (\nu-1)$, so does the $K_q$ term. The explicit form of the
Lommel function then yields
$$
\lim_{x\to0}\int x^{2\nu-q} K_{1+q}(x)=-2^{2\nu-q-1}(\nu-q-1)!\,\nu!,
\quad-1\le q
\le \nu-1.
\eql{0lim1}$$

When $q=\nu$, the inverse power terms in the series expansion of the Bessel
functions cancel on the right-hand side of (\peq{intq}) leaving just the
logarithm terms and positive powers which yield the limit\mgn{LOMMEL.MTH}
$$
\lim_{x\to0}\int x^\nu K_{1+\nu}(x)\sim-2^\nu\nu!\bigg(\ln(x/2)+\ga
-{1\over2}\sum_{k=1}^\nu{1\over k}\bigg),\quad \nu>0.
\eql{0lim2}$$
To regain (\peq{mint}) we set $x=2\pi mp$. As $m\to0$, (\peq{0lim2})
diverges
as $\ln m$ but this will be cancelled by other, irrelevant terms that we
do not need to calculate (except as a check).

Up to this log term, we find that the massless limit of the Bessel
terms obtained by substituting (\peq{watson10}) into (\peq{diffregsum})
and integrating with respect to $m^2$ is
$$\eqalign{
&{\nu!(\nu+1)!\over2\pi^{2\nu+2}}\,\ze_R(2\nu+2)\sum_{q=-1}^{\nu-1}
{\pi^{1/2}\over\Gamma(-q-1/2)}{(-1)^{\nu-q}\over(q+1)!(q-\nu)}\cr
&+{\nu!\pi^{1/2}\over\pi^{2\nu+2}\Gamma(-\nu-1/2)}\bigg(\big
(\ln\pi+\ga-{1\over2}\sum_{k=1}^\nu{1\over k}\big)\,
\ze_R(2\nu+2)-\ze_R'(2\nu+2)\bigg)\cr}
\eql{0lim3}$$
which, in view of the identities,
\begin{ignore}
$$\eqalign{
&{\nu!(\nu+1)!2^{2\nu}\over(2\nu+2)!}\,B_{2\nu+2}\sum_{q=-1}^{\nu-1}
{\sqrt\pi\over\Gamma(-q-1/2)}{(-1)^{\nu-q}\over(q+1)!(q-\nu)}\cr
&-{(-1)^\nu2^{2\nu+1}\sqrt\pi\nu!\over\Ga(2\nu+2)\Ga(-\nu-1/2)}
\ze_R'(-2\nu-1)\cr
&-{2^{2\nu+1}\nu!\sqrt\pi\over(2\nu+2)!\Ga(-\nu-1/2)}B_{2\nu+2}
\bigg(\ln2+{1\over2}\sum_{k=1}^\nu {1\over k}-\sum_{k=1}^
{2\nu+1}{1\over k}\bigg).\cr}
\eql{0lim4}$$
\end{ignore}
$$
(-1)^\nu{\nu!2^{2\nu+1}\sqrt\pi\over(2\nu+1)!\Ga(-\nu-1/2)}=-1.
$$
and\mgn{SUMID.MTH}
$$
{2^{2\nu+1}(\nu!)^2\over(2\nu+1)!}+\sum_{q=0}^{\nu-1}{2^{2\nu-2q}
\over\nu-q}\,\comb{2q+1}{q}\bigg/
{\comb{2\nu+1}\nu}=2\sum_{k=1}^{2\nu+1}{1\over k}-\sum_{k=1}^\nu{1\over k},
$$
simplifies considerably to
$$
\ze_R'(-2\nu-1)+{(-1)^\nu\over2(\nu+1)}B_{2\nu+2}\ln2
\eql{0lim5}$$
with no fractional part. The ultimate simplicity of this result suggests
the existence of an easier route.

When integrating (\peq{diffregsum}), the constant of integration necessary
to give a zero value for the regulated sum,
$\sumstar p p^{2\nu+1}\ln\big(2p/(p+\ep)\big)$, in the massless limit
is the negative of (\peq{0lim5}) and,
finally, this is the relevant term coming from (\peq{hsum}).

The remaining contribution in (\peq{logdet7}) is
$$
\sum_{p=1}^\infty p^{2\nu}\bigg(\sum_{n=1}^\infty{T_n(t)\over\ep^n}+
\int_0^\infty\bigg({1\over2}
-{1\over t}+{1\over e^t-1}\bigg){e^{-tp}\over t}\,dt\bigg).
\eql{rem}$$

The first thing to note is that for $n>2\nu+1$, the sum over $p$ of
$p^{2\nu}T_n/\ep^n$ converges and the result is irrelevant. This can be
shown by expanding the binomial $p^{2\nu}=\big(\ep^2-m^2\big)^{2\nu}$
and noting that, according to (\peq{as2}), only terms like $m^r/\ep^{r+1}$
sum to a  relevant part in the asymptotic limit. Alternatively, the limit,
[\pref{Moss}],\mgn{can set $p=1$ or can use $N_p$ argument}
$$
\sum_{p=0}^\infty{t^{2r}\over\ep^n}\sim{\Ga(r+1/2)\Ga\big((n-1)/2\big)
\over2\Ga(r+n/2)}m^{1-n},\quad r>0,\,n>1
\eql{Moss1}$$
can be used.

If the bracket in the integrand in (\peq{rem}) is completely expanded and
the integration performed, one gets the standard asymptotic formula,
$$\eqalign{
\int_0^\infty\bigg({1\over2}
-{1\over t}+{1\over e^t-1}\bigg){e^{-tp}\over t}\,dt&\sim
\sum_{k=1}^\infty(-1)^{k+1} {B_{2k}\over2k(2k-1)}{1\over p^{2k-1}}\cr
&=-\sum_{n=1}^\infty{T_n(1)\over p^n}\cr}
$$
where the last equality comes from Olver's relation between the $U_n(1)$
and
the coefficients in the Stirling approximation, [\pref{Olver}] (2.15),
(2.16). This shows that $T_n(1)$ vanishes when $n$ is even and that
$T_{2k-1}(1)=(-1)^kB_{2k}/2k(2k-1)$.

If the $n$ summation is restricted to the possibly relevant $n$-values
and appropriate terms
added to and subtracted from the integral, one finds for the possibly
relevant part of (\peq{rem})
$$\eqalign{
\sum_{p=1}^\infty p^{2\nu}\bigg[\sum_{n=1}^{2\nu+1}\bigg(&{T_n(t)
\over\ep^n}-{T_n(1)\over p^n}\bigg)\cr
&+\int_0^\infty\!\!\bigg({1\over2}
-{1\over t}+\!\sum_{k=1}^{\nu+1}(-1)^kB_{2k}{t^{2k-1}\over(2k)!}
+{1\over e^t-1}\bigg){e^{-tp}\over t}\,dt\bigg].\cr}
\eql{rem2}$$
The two components of this expression are looked at in turn. Consider
first
$$
\sum_{p=1}^\infty p^{2\nu}\sum_{n=1}^{2\nu+1}\bigg({T_n(t)\over\ep^n}-
{T_n(1)\over p^n}\bigg).
\eql{pol1}$$
The $T_n(t)$ are polynomials in $t$. Set
$$
T_n(t)=T_n(1)+T'_n(t).
$$
Hence one has for (\peq{pol1})
$$
\sum_{p=1}^\infty\sum_{n=1,3,}^{2\nu+1}T_n(1)
(\ep^2-m^2)^\nu\bigg({1\over\ep^n}-{1\over p^n}\bigg)
+\sum_{p=1}^\infty\sum_{n=1}^{2\nu+1}
{(\ep^2-m^2)^\nu\over\ep^n}T'_n(t).
\eql{pol2}$$

Most of the terms in the second summation can be dismissed as irrelevant
-- if they are divergent because of the $m$-dependence in $T'$ and if
convergent because of (\peq{as2}). The only relevant contribution is
when $n=2\nu+1$,
$$
\sum_{p=1}^\infty{t^{2\nu} T'_{2\nu+1}(t)\over\ep}.
\eql{sumd}$$

A convenient computational way of finding the asymptotic limit of
this sum is to extract a factor of $(1-t^2)=m^2/\ep^2$ from
$T_n'$, expand the resulting polynomial,
$$
T'_n(t)={m^2\over\ep^2}T''_n(t)={m^2\over\ep^2}\sum_{i=0}a_i^{(n)}t^{2i},
$$
and then  use formula (\peq{Moss1}) to get
$$\eqalign{
\sum_{p=1}^\infty{t^{2\nu} T'_{2\nu+1}(t)\over\ep}&=\sum_{i=0}
a_i^{(2\nu+1)}m^2\sum_{p=1}^\infty{t^{2i+2\nu}\over\ep^3}\sim\sum_{i=0}
{a_i^{(2\nu+1)}\over2i+2\nu+1}\cr
&=\int_0^1 t^{2\nu}T''_{2\nu+1}(t)\,dt.\cr}
\eql{polylim}$$

In the first summation in (\peq{pol2}) we can write, using (\peq{epexp}),
$$
\sum_{p=1}^\infty(\ep^2-m^2)^\nu\bigg({1\over\ep^n}-{1\over p^n}\bigg)\sim
\sum_{q=0}^\nu\comb\nu q (-1)^{\nu-q}m^{2\nu-2q}\sumstar p\big(\ep^{2q-n}
-p^{2q-n}\big).
\eql{1sum}$$

For nominally divergent terms to be relevant, there must be no factors of
$m$. (Refer to (\peq{watson10}).) This means that $\nu=q$ and then all
$n$ values are divergent. These yield\mgn{MandO's Bernoulli numbers}
$$\eqalign{
\sum_{n=1,3,\ldots}^{2\nu+1} T_n(1)\sumstar p
\big(\ep^{2\nu-n}-&p^{2\nu-n}\big)\sim\cr
&T_{2\nu+1}(1)\,(\ln2-\ga)-\sum_{k=1}^\nu T_{2k-1}(1)\,
\ze_R(2k-2\nu-1),\cr}
\eql{1sum1}$$
since the relevant part of $\sumstar p\big(\ep^{2\nu-n}-p^{2\nu-n}\big)$
is $-\ze_R(n-2\nu)$, for $n<2\nu+1$, according to (\peq{watson10}), and
$\ln2-\ga$ if $n=2\nu+1$ according to (\peq{bessum1}).

For $n=2\nu+1$ there are also convergent terms in (\peq{1sum}) which can
be evaluated using (\peq{as2}) to give a relevant contribution,
$$\eqalign{
T_{2\nu+1}(1)\sum_{q=1}^\nu&\comb\nu q(-1)^q\sum_p{m^{2q}
\over\ep^{2q+1}}\cr
&\sim T_{2\nu+1}(1)\sum_{q=1}^\nu
\comb\nu q{(-1)^q\sqrt\pi\Ga(q)\over2\Ga(q+1/2)}.\cr}
\eql{1sum2}
$$
The combination of (\peq{1sum1}) and (\peq{1sum2}) is
$$\eqalign{
-\sum_{k=1}^\nu&{(-1)^k B_{2k}\over2k(2k-1)}\ze_R(2k-2\nu-1)\cr
&+T_{2\nu+1}(1)\bigg(\ln2-\ga
+{1\over2}\sum_{q=1}^\nu{(-1)^q\sqrt\pi\nu!\over q(\nu-q)!
\Ga(q+1/2)}\bigg).\cr}
\eql{total1}$$

The integral in (\peq{rem2}) remains to be evaluated. Performing the sum
over $p$ and introducing a regularisation, it becomes
$$
\lim_{s\to0}\int_0^\infty\bigg({1\over2}
-{1\over t}+\sum_{k=1}^{\nu+1}(-1)^kB_{2k}{t^{2k-1}\over(2k)!}
+{1\over e^t-1}\bigg)t^{s-1}{d^{2\nu}\over dt^{2\nu}}{1\over e^t-1}\,dt.
\eql{int3}$$

The power terms have been dealt with earlier in (\peq{powers1}).
The remaining part is
$$
\lim_{s\to0}\int_0^\infty{1\over e^t-1}t^{s-1}
{d^{2\nu}\over dt^{2\nu}}{1\over e^t-1}\,dt.
\eql{dint}$$

Writing
$$
(-1)^j{d^j\over dt^j}{1\over e^t-1}=\sum_{l=1}^{j+1}{A_l^{(j)}\over
(e^t-1)^l}
$$with\mgn{EXPCOEFS.MTH}
$$\eqalign{
&A_l^{(j)}=l A_l^{(j-1)}+(l-1)A_{l-1}^{(j-1)},\quad 2\le l\le j\cr
&A_{j+1}^{(j)}=j!\cr
&A_1^{(j)}=1\cr}
\eql{recurs}$$(which is easily iterated by machine) we find for
(\peq{dint})
$$
\lim_{s\to0}\sum_{l=1}^{2\nu+1}A_l^{(2\nu)}\,\Ga(s)\,\ze_{l+1}(s,l+1)
\eql{dint2}$$
the finite part of which is
$$
\sum_lA_l^{(2\nu)}\big(\ze'_{l+1}(0,l+1)-\ga\ze_{l+1}(0,l+1)\big).
\big.
\eql{finint2}$$

Equation (\peq{barnexp}) allows the Barnes \zfs\ to be rewritten in terms
of the Riemann \zf. Combining the two parts of the integral term gives
$$\eqalign{
\lim_{s\to0}\bigg({1\over2}\Ga(s)&\ze_R(s-2\nu)-\Ga(s-1)\ze_R(s-2\nu-1)\cr
&-\sum_{k=1}^{\nu+1}(-1)^{k+1}{B_{2k}\over(2k)!}\Ga(s+2k-1)\ze_R(s+2k
-2\nu-1)\cr
&\hspace{*****}+\Ga(s)\sum_{l=1}^{2\nu+1}\sum_{k=1}^{l+1}\,A_l^{(2\nu)}
\,{S^{(k)}_{l+1}\over l!}\,\ze_R(s-k+1) \bigg).\cr}
\eql{total2}$$

The cancellation of the individual divergences is a check of the analysis
and implies the identity
$$
\ze_R(-2\nu-1)-{(-1)^\nu B_{2\nu+2}\over(2\nu+2)!}\Ga(2\nu+1)
+\sum_{l=1}^{2\nu+1}
\sum_{k=1}^{l+1}\,A_l^{(2\nu)}\,{S^{(k)}_{l+1}\over l!}\,\ze_R(-k+1)=0,
\eql{polcanc}$$
where $k$ will actually be even.\mgn{$k=1$ goes out because of sumrule}

The finite part is
$$\eqalign{
{1\over2}\ze_R'(&-2\nu)+\ze_R'(-2\nu-1)-(\ga-1)\ze_R(-2\nu-1)
-\ga{(-1)^\nu B_{2\nu+2}\over(2\nu+2)!}\Ga(2\nu+1)\cr
&-\sum_{k=1}^\nu(-1)^{k+1}{B_{2k}\over(2k)!}\Ga(2k-1)\ze_R(2k-2\nu-1)
+T_{2\nu+1}(1)\psi(2\nu+1)\cr
&\hspace{*****}+\sum_{l=1}^{2\nu+1}\sum_{k=1}^{l+1}\,A_l^{(2\nu)}\,
{S^{(k)}_{l+1}\over l!}\,\big(\ze'_R(-k+1)-\ga\ze_R(-k+1)\big)\cr}
$$
which can be reduced using (\peq{polcanc}) leaving,
$$\eqalign{
{1\over2}&\ze_R'(-2\nu)+\ze_R'(-2\nu-1)+\ze_R(-2\nu-1)
+T_{2\nu+1}(1)\big(\psi(2\nu+1)+2\ga\big)\cr
&-\sum_{k=1}^\nu(-1)^{k+1}{B_{2k}\over(2k)!}\Ga(2k-1)\,\ze_R(2k-2\nu-1)
+\sum_{l=1}^{2\nu+1}\sum_{k=1}^{l+1}\,A_l^{(2\nu)}\,{S^{(k)}_{l+1}\over l!}
\,\ze'_R(-k+1).\cr}
\eql{total3}$$

A certain amount of simplification occurs if (\peq{total3}) is combined
with
(\peq{total1}). The first summation in (\peq{total3}) goes out and a $\ga$
term cancels so that the $\psi$ can combine with the last $\ga$ term to
remove any remaining $\ga$-dependence. Including also the relevant parts
of $\sumstar{p}p^{2\nu}(\ep-p)$ and $-{1\over2}\sumstar{p} p^{2\nu}
\ln(\ep/p)$ obtained earlier, we find
$$\eqalign{
\ze_R'(-2\nu-1)&+\sum_{l=1}^{2\nu+1}\sum_{k=1}^{l+1}\,A_l^{(2\nu)}\,
{S^{(k)}_{l+1}\over l!}\,\ze'_R(-k+1)\cr
&+T_{2\nu+1}(1)\bigg(\ln2
+\sum_{k=1}^{2\nu}{1\over k}+\sum_{q=1}^\nu{(-1)^q
\sqrt\pi\nu!\over2q(\nu-q)!\Ga(q+1/2)}\bigg).\cr}
\eql{total5}$$

Added to (\peq{sumd}) and (\peq{0lim5}), this would be the final answer
for (\peq{logdet7}), but first we manipulate (\peq{total5}) a little.
Specifically, the order of summation is changed,
$$\eqalign{
\sum_{l=1}^{2\nu+1}&\sum_{k=1}^{l+1}\,A_l^{(2\nu)}\,
{S^{(k)}_{l+1}\over l!}\,\ze'_R(-k+1)\cr
&=\sum_{k=1}^{2\nu}
M^{(2\nu)}_k\,\ze'_R(-k)
+{1\over2\nu+1}\ze_R'(-2\nu-1)=\sum_{k=1}^{2\nu+1}
M^{(2\nu)}_k\,\ze'_R(-k)\cr}
$$
\begin{ignore}
$$\eqalign{
\sum_{l=1}^{2\nu+1}&\sum_{k=1}^{l+1}\,A_l^{(2\nu)}\,
{S^{(k)}_{l+1}\over l!}\,\ze'_R(-k+1)\cr
&=\sum_{k=1}^{2\nu+1}\sum_{l=k}^{2\nu+1}\,A_l^{(2\nu)}\,
{S^{(k)}_{l+1}\over l!}\,\ze'_R(-k+1)+\sum_{l=1}^{2\nu+1}
{A^{(2\nu)}_l\over l!}\ze_R'(-l)\cr
&=\sum_{k=0}^{2\nu}\sum_{l=k+1}^{2\nu+1}\,A_l^{(2\nu)}\,
{S^{(k+1)}_{l+1}\over l!}\,\ze'_R(-k)+\sum_{l=1}^{2\nu+1}
{A^{(2\nu)}_l\over l!}\ze_R'(-l)\cr
&=\sum_{k=1}^{2\nu}\sum_{l=k+1}^{2\nu+1}\,A_l^{(2\nu)}\,
{S^{(k+1)}_{l+1}\over l!}\,\ze'_R(-k)+\sum_{k=1}^{2\nu+1}
{A^{(2\nu)}_k\over k!}\ze_R'(-k)\cr
&=\sum_{k=1}^{2\nu}\bigg({A^{(2\nu)}_k\over k!}+
\sum_{l=k+1}^{2\nu+1}\,A_l^{(2\nu)}\,
{S^{(k+1)}_{l+1}\over l!}\bigg)\,\ze'_R(-k)
+{1\over2\nu+1}\ze_R'(-2\nu-1)\cr
&=\sum_{k=1}^{2\nu}
\sum_{l=k}^{2\nu+1}\,A_l^{(2\nu)}\,
{S^{(k+1)}_{l+1}\over l!}\,\ze'_R(-k)
+{1\over2\nu+1}\ze_R'(-2\nu-1)\cr
&=\sum_{k=1}^{2\nu}
M^{(2\nu)}_k\,\ze'_R(-k)
+{1\over2\nu+1}\ze_R'(-2\nu-1)=\sum_{k=1}^{2\nu+1}
M^{(2\nu)}_k\,\ze'_R(-k)\cr}
$$
\end{ignore}
where
$$
M^{(2\nu)}_k\equiv
\sum_{l=k}^{2\nu+1}\,A_l^{(2\nu)}\,{S^{(k+1)}_{l+1}\over l!}.
$$
We have used the Stirling number value $S^{(r)}_r=1$. We have also
used the values $S^{(1)}_{l+1}=(-1)^ll!$, and the sum rule,
$$
\sum_{l=1}^{2\nu+1}(-1)^lA^{(2\nu)}_l=0,
$$
(which is easily proved from the recursion relation (\peq{recurs})) to
alter
a lower summation limit from $k=0$ to $k=1$ in the intermediate algebra.

(\peq{total5}) now becomes \mgn{CHECK SIGNS}
$$\eqalign{
\ze'_R(-2\nu-1)&+\sum_{k=1}^{2\nu+1}M^{(2\nu)}_k\,\ze'_R(-k)\cr
&+T_{2\nu+1}(1)\bigg(\ln2
+\sum_{k=1}^{2\nu}{1\over k}+\sum_{q=1}^\nu
{(-1)^q\sqrt\pi\nu!\over2q(\nu-q)!\Ga(q+1/2)}\bigg).\cr}
\eql{total6}$$

The final contribution to (\peq{logdet7}) is the asymptotic form
(\peq{polylim}). Adding this to (\peq{total6}) and subtracting the Bessel
function contribution (\peq{0lim5}), we obtain the complete Dirichlet
expression, \mgn{FRACCONS.MTH}
$$\eqalign{
\sum_{k=1}^{2\nu+1}M^{(2\nu)}_k\,&\ze'_R(-k)-{(-1)^\nu B_{2\nu+2}\over
2\nu+1}\ln2+\int_0^1t^{2\nu}T''_{2\nu+1}(t)\,dt\cr
&+T_{2\nu+1}(1)\bigg(\sum_{k=1}^{2\nu}
{1\over k}+\sum_{q=1}^\nu
{(-1)^q\sqrt\pi\nu!\over2q(\nu-q)!\Ga(q+1/2)}\bigg).\cr}
\eql{total7}$$

The particular values
$$
A^{(j)}_j={1\over2}(j+1)!,\quad S^{(r-1)}_r=-{1\over2}r(r-1)
$$
are readily obtained and give
$$
M^{(2\nu)}_{2\nu}=-{1\over2}.
$$
Also $M^{(2\nu)}_{2\nu+1}=1/(2\nu+1)$.

The final step to finding $\ze'(0)$ on an even ball is to compound
(\peq{total7}) with the expansion of the degeneracy (\peq{degene}),
$$
N_p^{(d)}={2\over(d-2)!}\sum_{\nu=1}^r S^{(\nu)}_r p^{2\nu},\quad
r=1+(d/2-2)^2.
\eql{expdeg}$$

The explicit evaluation for any specific dimension $d$ is straightforward,
all but the rational part being quickly done. The integral term in
(\peq{total7}) is the
hardest to find since it involves the cumulant expansion, (\peq{cumul}).

The examples that have been calculated agree with those exhibited by Bordag
{\it et al} [\pref{BGKE}] and in [\pref{DandA}] for $d=4$.
The $d=6$ expression is repeated here for completeness,
$$\eqalign{
\ze'_{6\rm ball}(0)&=-{4027\over6486480}-{1\over756}\ln2+{1\over60}
\ze_R'(-5)\cr
&\hspace{***}-{1\over24}\ze_R'(-4)+{1\over24}\ze_R'(-2)-{1\over60}\ze_R'(-1)\cr
\noalign{\vskip5truept}
&\approx\,-0.000392.\cr}
$$

We also give the coefficients of the $\ze_R'(-k)$ terms for $d=8$
in the form of a vector labelled by $k$,
$$
\bigg[-{1\over135},-{7\over144},{79\over2160},{5\over72},-{31\over648},
-{1\over30},{19\over990}\bigg].
$$

\section {\bf 9. Robin conditions.}

The Robin case proceeds in very similar fashion. The starting point is the
expression

$$\eqalign{
\ln\De\big(-m^2\big)\sim&
\sum_{p=1}^\infty p^{2\nu}\bigg[p\ln{2p\over p+\ep}+\ep-p
+{1\over2}\ln{\ep\over p}-\ln(1+\be/p)+\sum_{n=1}^\infty{R_n(\be,t)\over
\ep^n}\cr
&\hspace{***}-{\rm Ray}(\be,m,p)+\int_0^\infty\bigg({1\over2}
-{1\over t}+{1\over e^t-1}\bigg){e^{-tp}\over t}\,dt\bigg].\cr}
\eql{logdetr1}$$

Again, so far as relevant terms go, the Rayleigh contribution can be
ignored, as can the $n>2\nu+1$ terms in the summation.

{}From the formulae in section 3, it is not difficult to derive the relation
$$
\ln\bigg(\sum_{n=0}^\infty{\overline W_n(1)\over p^n}\bigg)
=\ln\big(1+\be/p\big)+\ln\bigg(\sum_{n=0}^\infty{\overline U_n(1)\over p^n}
\bigg).
$$which implies (\cf [\pref{BGKE}])
$$
R_n(\be,1)=T_n(1)+{(-1)^{n+1}\over n}\be^n.
$$

This allows one to add and subtract terms to rewrite the effective part of
(\peq{logdetr1}) as
$$\eqalign{
\ln\De\big(-m^2\big)&\sim
\sumstar{p}p^{2\nu}\bigg(p\ln{2p\over p+\ep}+\ep-p
+{1\over2}\ln{\ep\over p}-\ln(1+\be/p)\bigg)\cr
&+\sum_{p=1}^\infty p^{2\nu}\bigg[\sum_{n=1}^{2\nu+1}{R_n(\be,t)-R_n(\be,1)
\over\ep^n}+\sum_{n=1}^{2\nu+1}R_n(\be,1)\bigg({1\over\ep^n}-
{1\over p^n}\bigg)\cr
&\hspace{*}+\int_0^\infty\bigg({1\over2}
-{1\over t}-\!\sum_{k=1}^{\nu+1}(-1)^{k+1}B_{2k}{t^{2k-1}\over(2k)!}
+{1\over e^t-1}\bigg){e^{-tp}\over t}\,dt\bigg].\cr}
\eql{logdetr2}$$

The only really new term is $\sumstar{p}p^{2\nu}\ln(1+\be/p)$ which is
easily reduced to
$$
\sumstar{p}p^{2\nu}\ln(1+\be/p)=-{\ga\over2\nu+1}\be^{2\nu+1}
-\int_0^\be\be^{2\nu}\psi(1+\be)\,d\be.
\eql{exterm}$$

Most of the other terms are identical to the Dirichlet case but the
cancellations are not as complete and we find
for the total expression corresponding to (\peq{total7}),
$$\eqalign{
\ze_R&'(-2\nu)+\sum_{k=1}^{2\nu+1}M^{(2\nu)}_k\,\ze'_R(-k)+{1\over2\nu+1}
\big(\be^{2\nu+1}-(-1)^\nu B_{2\nu+2}\big)\ln2\cr
&+\int_0^\be\be^{2\nu}\psi(1+\be)\,d\be+\int_0^1 t^{2\nu}R''_{2\nu+1}(\be,t)
\,dt+T_{2\nu+1}(1)\sum_{k=1}^{2\nu}{1\over k}-{1\over4\nu}\be^{2\nu} \cr
&-\sum_{k=1}^\nu{\be^{2k-1}\over2k-1}\,\ze_R(2k-2\nu-1)+\big(T_{2\nu+1}(1)
+{\be^{2\nu+1}\over2\nu+1}\big)\sum_{q=1}
^\nu{(-1)^q\nu!\sqrt\pi\over2q(\nu-q)!\Ga(q+1/2)}.\cr}
\eql{totalr}$$
Again this has to be compounded with the expansion of the degeneracy.

For $d=6$ we find
$$\eqalign{
\ze_6'(0)=&-{9479\over32432400}-{11\over315}\be+{517\over15120}\be^2
+{41\over1512}\be^3-{3\over160}\be^4-{1\over45}\be^5\cr
&\hspace{*}-\bigg[{1\over756}+{1\over180}\be^3\big(5-3\be^2\big)\bigg]\ln2
+{1\over12}\int_0^1\be^2\big(\be^2-1\big)\psi(1+\be)\,d\be\cr
&\hspace{**}+{1\over60}
\ze_R'(-5)+{1\over24}\ze_R'(-4)-{1\over24}\ze_R'(-2)-{1\over60}\ze_R'(-1)\cr}
$$
again agreeing with [\pref{BGKE}] after some rearrangement.

The coefficients of the $\ze_R'(-k)$ for $d=8$ are
$$
\bigg[-{1\over135},{7\over144},{79\over2160},-{5\over72},-{31\over648},
{1\over30},{19\over990}\bigg].
$$
\section{\bf 10. The massive case.}
Although our main interest is with massless fields, we indicate in this section
a means of computing the finite mass determinant, $\ln D(-m^2)=-\ze'(0,m^2)$.
This will follow from the basic formula (\peq{vreln}) which is repeated here
$$\eqalign{
\ze'(0,m^2)=&\ze'(0,0)-\ln\De(-m^2)\cr
&+FP\ze(k,0){z^{2k}\over k}-\sum_{k=2}^{[\mu]}C_{d/2-k}\big(1+\ldots
+{1\over k+1}\big){z^{2k}\over k!}.\cr}
\eql{vreln2}$$
$\ln\De(-m^2)$ can be found from the properties of the Bessel functions
but the finite part
$$\eqalign{
FP\ze(k,0)&\equiv\ze(k,0)\quad(k\,\,{\rm not\,\,a\,\,pole)}\cr
&\equiv\lim_{\ep\to0}\bigg(\ze(k+\ep,0)-{R\over\ep}\bigg)
\quad(k\,\,{\rm is\,\,a\,\,pole)},\cr}
\eql{finp}$$
is a transcendental quantity and its direct computation would involve an
analytic continuation of the \zf. However, since we know, in principle, the
precise large-$m$ dependence of the other quantities in (\peq{vreln2}),
$FP\ze(k,0)$ can simply be read off. This procedure is now illustrated for
the disc, $[\mu]=1$.

{}From the known asymptotic behaviour of $\ze'(0,m^2)$ in terms of the
heat-kernel coefficients, and from the $m^2$ dependence of the quantities in
(\peq{logdet2}) obtained in section 5, one can confirm that the only term in
$\ln\big(\De(-m^2)/D(-m^2)\big)$ that  does not cancel is proportional to
$m^2$. Therefore, since $\ze(s,0)$ has a pole at $s=1$ of residue $1/4$,
$$
\lim_{s\to1}\bigg(\ze(s,0)-{1\over4(s-1)}\bigg)={1\over2}(\ga-1-\ln2).
\eql{remain1}$$

On the disc, inserting the power series for $I_p$ into (\peq{2disc}),
$$
\ln\De(-m^2)=\sum_{p=0}^\infty N_p^{(2)}\bigg[\ln\bigg(1+\sum_{k=1}^\infty
{p!\over k!(p+k)!}\big({m\over k}\big)^{2k}\bigg)-{m^2\over4(1+p)}\bigg],
\eql{mdet1}$$
with the extension to higher dimensions being obvious.

The Rayleigh terms can be found from the Taylor series (cumulant) expansion of
the logarithm and, making this expansion, one obtains a power series in $m^2$
that begins with the first term not subtracted. Then, in terms of the
well-known Rayleigh functions defined by
$$
\si^{(l)}(p)=\sum_{\al_p}{1\over\al_p^{2l}},
$$
(\peq{mdet1}) becomes
$$
\ln\De(-m^2)=\sum_{l=2}^\infty\sum_{p=0}^\infty N_p^{(2)}\si^{(l)}(p)\,m^{2l}
=\sum_{l=2}\ze(l,0)\,m^{2l}.
$$
In $d$-dimensions, the $l$-sum starts at $[d/2]+1$.

The expressions for the Rayleigh functions allow the sum over $p$ to be
performed. The first few terms are
$$\eqalign{
\ze'(0,m^2)&=\ze'(0,0)-{1\over2}(\ga-1-\ln2)m^2-{1\over192}(2\pi^2-15)m^4\cr
&\hspace{*************}-{1\over3072}\big(35-6\pi^2+24\ze_R(3)\big)m^6
+\ldots\cr
&\approx\ze'(0,0)+0.557966m^2-0.0246834m^4+0.0020103m^6+\ldots\cr}
\eql{mdet2})$$

Of course, this series follows more directly by expanding the definition
(\peq{zem}) but it would then be necessary to find the remainder
(\peq{remain1}) by an independent calculation.

For comparison, the large positive $m$ expansion is
$$
\ze'(0,m^2)\sim{1\over4}m^2(\ln m^2-1)+{\pi\over2}m-{1\over6}\ln m^2+
{\pi\over128m}+{2\over315m^2}+{37\pi\over2^{14}m^3}+\ldots
\eql{largem}$$
and a  graphical analysis shows very good agreement between this and
(\peq{mdet2}) for $m^2$ in the range $0.6$ to $0.8$. We should remark that
Elizalde [\pref{Elizalde1}] has performed numerical calculations on similar
expansions.
\section{\bf 11. Comments}
In this paper, which is a mainly technical one, we have presented methods
for the evaluation of the
functional determinants on the $d$-ball for Dirichlet and Robin boundary
conditions applied to a massless scalar field.

The information required is the asymptotic behaviour of the
eigenfunctions, the eigenvalue properties being introduced via the
Mittag-Leffler theorem. The method involves extracting the
mass-independent terms from a large-mass asymptotic limit.

For the disc and 4-ball, the calculation is done in more detail than
required. In these cases, complete expansions are obtained which allow the
necessary cancellations of the mass-dependent terms to be confirmed. The
particular summations encountered in these explicit evaluations may have
independent interest.

For arbitrary dimension (actually only even here) a systematic procedure
has been presented and an explicit formula given for the functional
determinant in terms of coefficients evaluated by recursion.

The method of [\pref{BEK}] has been also used by  Bordag, Geyer, Kirsten
and Elizalde [\pref{BGKE}] recently to evaluate the functional
determinants on balls directly from properties of Bessel functions in an
effective and systematic fashion. There are many similarities in the
intermediate expressions to our work but the details are different.
They give precise results as far as $d=6$ and indicate a general method.

The spin-1/2 values are available for $d=2,3$ and 4 (J.S.Apps) by
conformal methods and one anticipates only technical differences in
extending the present approach to this case.

The corresponding calculation on spherical caps should also be possible.
(See Moss [\pref{Moss}] and Barvinsky {\it et al} [\pref{Barv}].)

\section{\bf Acknowledgments}
I would like to thank Klaus Kirsten for communicating his methods and
results as well as for help. Thanks also to Emilio Elizalde for
information on his approach to asymptotic expansions.

\begin{ignore}
$$\eqalign{
\sum_p{K_0(2\pi mp)\over p^2}=&\sum_n
{1\over24}\bigg((2(3b^2+6\pi b+2\pi^2)-3a)
\ln\big(\sqrt(a+b^2+4\pi b+4\pi^2)+b+2\pi\big)\cr
&+(3a-2(3b^2+6\pi b+2\pi^2))\ln\big(\sqrt(a+b^2)+b\big)\cr
&-3(3b+2\pi)\big(\sqrt(a+b^2+4\pi b+4\pi^2)-
\sqrt(a+b^2)\big)\bigg)\cr}
$$
(2*(3*b^2+6*pi*b+2*pi^2)-3*a)*LOG(SQR(a+b^2+4*pi*b+4*pi^2)+b+2*pi)/24+
(3*a-2*(3*b^2+6*pi*b+2*pi^2))*LOG(SQR(a+b^2)+b)/24
-(SQR(a+b^2+4*pi*b+4*pi^2)*(3*b+2*pi)-SQR(a+b^2)*(3*b+4*pi))/8
\end{ignore}

\vskip 10truept
\noin{\bf{References}}
\vskip 5truept
\begin{putreferences}
\ref{Rayleigh}{Lord Rayleigh{\it Theory of Sound} vols.I and II,
MacMillan, London, 1877,78.}
\ref{KCD}{G.Kennedy, R.Critchley and J.S.Dowker \aop{125}{80}{346}.}
\ref{Donnelly}{H.Donnelly \ma{224}{1976}161.}
\ref{Fur2}{D.V.Fursaev {\sl Spectral geometry and one-loop divergences on
manifolds with conical singularities}, JINR preprint DSF-13/94,
hep-th/9405143.}
\ref{HandE}{S.W.Hawking and G.F.R.Ellis {\sl The large scale structure of
space-time} Cambridge University Press, 1973.}
\ref{DandK}{J.S.Dowker and G.Kennedy \jpa{11}{78}{895}.}
\ref{ChandD}{Peter Chang and J.S.Dowker \np{395}{93}{407}.}
\ref{FandM}{D.V.Fursaev and G.Miele \pr{D49}{94}{987}.}
\ref{Dowkerccs}{J.S.Dowker \cqg{4}{87}{L157}.}
\ref{BandH}{J.Br\"uning and E.Heintze \dmj{51}{84}{959}.}
\ref{Cheeger}{J.Cheeger \jdg{18}{83}{575}.}
\ref{SandW}{K.Stewartson and R.T.Waechter \pcps{69}{71}{353}.}
\ref{CandJ}{H.S.Carslaw and J.C.Jaeger {\it The conduction of heat
in solids} Oxford, The Clarendon Press, 1959.}
\ref{BandH}{H.P.Baltes and E.M.Hilf {\it Spectra of finite systems}.}
\ref{Epstein}{P.Epstein \ma{56}{1903}{615}.}
\ref{Kennedy1}{G.Kennedy \pr{D23}{81}{2884}.}
\ref{Kennedy2}{G.Kennedy PhD thesis, Manchester (1978).}
\ref{Kennedy3}{G.Kennedy \jpa{11}{78}{L173}.}
\ref{Luscher}{M.L\"uscher, K.Symanzik and P.Weiss \np {173}{80}{365}.}
\ref{Polyakov}{A.M.Polyakov \pl {103}{81}{207}.}
\ref{Bukhb}{L.Bukhbinder, V.P.Gusynin and P.I.Fomin {\it Sov. J. Nucl.
 Phys.} {\bf 44} (1986) 534.}
\ref{Alvarez}{O.Alvarez \np {216}{83}{125}.}
\ref{DandS}{J.S.Dowker and J.P.Schofield \jmp{31}{90}{808}.}
\ref{Dow1}{J.S.Dowker \cmp{162}{94}{633}.}
\ref{Dow2}{J.S.Dowker \cqg{11}{94}{557}.}
\ref{Dow3}{J.S.Dowker \jmp{35}{94}{4989}; erratum {\it ibid}, Feb.1995.}
\ref{Dow5}{J.S.Dowker {\it Heat-kernels and polytopes} To be published}
\ref{Dow6}{J.S.Dowker \pr{D50}{94}{6369}.}
\ref{Dow7}{J.S.Dowker \pr{D39}{89}{1235}.}
\ref{BandG}{P.B.Gilkey and T.P.Branson \tams{344}{94}{479}.}
\ref{Schofield}{J.P.Schofield Ph.D.thesis, University of Manchester,
(1991).}
\ref{Barnesa}{E.W.Barnes {\it Trans. Camb. Phil. Soc.} {\bf 19} (1903)
374.}
\ref{BandG2}{T.P.Branson and P.B.Gilkey {\it Comm. Partial Diff. Equations}
{\bf 15} (1990) 245.}
\ref{Pathria}{R.K.Pathria {\it Suppl.Nuovo Cim.} {\bf 4} (1966) 276.}
\ref{Baltes}{H.P.Baltes \prA{6}{72}{2252}.}
\ref{Spivak}{M.Spivak {\it Differential Geometry} vols III, IV, Publish
or Perish, Boston, 1975.}
\ref{Eisenhart}{L.P.Eisenhart {\it Differential Geometry}, Princeton
University Press, Princeton, 1926.}
\ref{Moss}{I.Moss \cqg{6}{89}{659}.}
\ref{Barv}{A.O.Barvinsky, Yu.A.Kamenshchik and I.P.Karmazin \aop {219}
{92}{201}.}
\ref{Kam}{Yu.A.Kamenshchik and I.V.Mishakov {\it Int. J. Mod. Phys.}
{\bf A7} (1992) 3265.}
\ref{DandE}{P.D.D'Eath and G.V.M.Esposito \prD{43}{91}{3234}.}
\ref{Rich}{K.Richardson \jfa{122}{94}{52}.}
\ref{Osgood}{B.Osgood, R.Phillips and P.Sarnak \jfa{80}{88}{148}.}
\ref{BCY}{T.P.Branson, S.-Y. A.Chang and P.C.Yang \cmp{149}{92}{241}.}
\ref{Vass}{D.V.Vassilevich.{\it Vector fields on a disk with mixed
boundary conditions} gr-qc /9404052.}
\ref{MandP}{I.Moss and S.Poletti \pl{B333}{94}{326}.}
\ref{Kam2}{G.Esposito, A.Y.Kamenshchik, I.V.Mishakov and G.Pollifrone
\prD{50}{94}{6329}.}
\ref{Aurell1}{E.Aurell and P.Salomonson \cmp{165}{94}{233}.}
\ref{Aurell2}{E.Aurell and P.Salomonson {\it Further results on functional
determinants of laplacians on simplicial complexes} hep-th/9405140.}
\ref{BandO}{T.P.Branson and B.\O rsted \pams{113}{91}{669}.}
\ref{Elizalde1}{E.Elizalde\jmp{36}{94}{3308}.}
\ref{BandK}{M.Bordag and K.Kirsten {\it Heat-kernel coefficients of
the Laplace operator on the 3-dimensional ball} hep-th/9501064.}
\ref{Waechter}{R.T.Waechter \pcps{72}{72}{439}.}
\ref{GRV}{S.Guraswamy, S.G.Rajeev and P.Vitale {\it O(N) sigma-model as
a three dimensional conformal field theory}, Rochester preprint UR-1357.}
\ref{CandC}{A.Capelli and A.Costa \np {314}{89}{707}.}
\ref{IandZ}{C.Itzykson and J.-B.Zuber \np{275}{86}{580}.}
\ref{BandH}{M.V.Berry and C.J.Howls \prs {447}{94}{527}.}
\ref{DandW}{A.Dettki and A.Wipf \np{377}{92}{252}.}
\ref{Weisbergerb} {W.I.Weisberger \cmp{112}{87}{633}.}
\ref{Voros}{A.Voros \cmp{110}{87}{110}.}
\ref{Pockels}{F.Pockels {\it \"Uber die partielle Differentialgleichung
$\Delta u+k^2u=0$}, B.G.Teubner, Leipzig 1891.}
\ref{Kober}{H.Kober \mz{39}{1935}{609}.}
\ref{Watson2}{G.N.Watson \qjm{2}{31}{300}.}
\ref{DandC1}{J.S.Dowker and R.Critchley \prD {13}{76}{3224}.}
\ref{Lamb}{H.Lamb \pm{15}{1884}{270}.}
\ref{EandR}{E.Elizalde and A.Romeo International J. of Math. and Phys.
{\bf13} (1994) 453}
\ref{DandA}{J.S.Dowker and J.S.Apps \cqg{12}{95}{1363}.}
\ref{Watson1}{G.N.Watson {\it Theory of Bessel Functions} Cambridge
University Press, Cambridge, 1944.}
\ref{BGKE}{M.Bordag, B.Geyer, K.Kirsten and E.Elizalde, {\it Zeta function
determinant of the Laplace operator on the D-dimensional ball} UB-ECM-PF
95/10, hep-th/9505157.}
\ref{MandO}{W.Magnus and F.Oberhettinger {\it Formeln und S\"atze}
Springer-Verlag, Berlin, 1948.}
\ref{Olver}{F.W.J.Olver {\it Phil.Trans.Roy.Soc} {\bf A247} (1954) 328.}
\ref{Hurt}{N.E.Hurt {\it Geometric Quantization in action} Reidel,
Dordrecht, 1983.}
\ref{Esposito}{G.Esposito {\it Quantum Gravity, Quantum Cosmology and
Lorentzian Geometry}, Lecture Notes in Physics, Monographs, Vol. m12,
Springer-Verlag, Berlin 1994.}
\ref{Louko}{J.Louko \prD{38}{88}{478}.}
\ref{Schleich} {K.Schleich \prD{32}{85}{1989}.}
\ref{BEK}{M.Bordag, E.Elizalde and K.Kirsten {\it Heat kernel
coefficients of
the Laplace operator on the D-dimensional ball} UB-ECM-PF 95/3.}
\ref{ELZ}{E.Elizalde, S.Leseduarte and S.Zerbini. hep-th/9303126.}
\ref{BGV}{T.P.Branson, P.B.Gilkey and D.V.Vassilevich {\it The Asymptotics
of the Laplacian on a manifold with boundary} II, hep-th/9504029.}
\ref{Erdelyi}{A.Erdelyi,W.Magnus,F.Oberhettinger and F.G.Tricomi {\it
Higher Transcendental Functions} Vol.I McGraw-Hill, New York, 1953.}
\ref{Quine}{J.R.Quine, S.H.Heydari and R.Y.Song \tams{338}{93}{213}.}
\ref{Dikii}{L.A.Dikii {\it Usp. Mat. Nauk.} {\bf13} (1958) 111.}
\end{putreferences}
\bye